\documentclass[acmtog, nonacm]{acmart}
\usepackage{tikz}
\usepackage{booktabs}
\usepackage{tabularx} 
\usepackage{array}     
\usepackage{caption}   
\usepackage{xcolor}   
\usepackage{bm,microtype}

\usepackage[noend]{algpseudocode}

\usepackage{mathrsfs}
\usepackage{svg}
\usepackage{xcolor}
\usepackage{wrapfig}
\usepackage[export]{adjustbox}
\usepackage{colortbl}
\usepackage[switch]{lineno}
\usepackage{mdframed}
\usepackage[ruled,linesnumbered]{algorithm2e} 

\SetCommentSty{mycommfont}
\let\oldnl\nl

\SetAlFnt{\small}
\SetAlCapFnt{\small}
\SetAlCapNameFnt{\small}
\SetAlCapHSkip{0pt}
\newcommand{\nonl}{\renewcommand{\nl}{\let\nl\oldnl}}
\newlength\savedwidth

\AtBeginDocument{%
  }

\acmSubmissionID{116}

\begin{document}

\title{M-ABD: Scalable, Efficient, and Robust Multi-Affine-Body Dynamics}


\author{Zhiyong He}
\affiliation{%
  \institution{University of Utah}
  \city{Salt Lake City}
  \country{USA}}
\email{zhiyong.he@utah.edu}

\author{Dewen Guo}
\affiliation{%
  \institution{University of Utah}
  \city{Salt Lake City}
  \country{USA}}
\email{guodw.sh@gmail.com}

\author{Minghao Guo}
\affiliation{%
  \institution{MIT}
  \city{Cambridge}
  \country{USA}}
\email{guomh2014@gmail.com}

\author{Yili Zhao}
\affiliation{%
  \institution{University of Southern California}
  \city{Los Angeles}
  \country{USA}}
\email{yilizhao.cs@gmail.com}

\author{Wojciech Matusik}
\affiliation{%
  \institution{MIT}
  \city{Cambridge}
  \country{USA}}
\email{wojciech@mit.edu}

\author{Hao Su}
\affiliation{%
  \institution{UCSD}
  \city{San Diego}
  \country{USA}}
\email{academic@haosu.ai}

\author{Chenfanfu Jiang}
\affiliation{%
  \institution{UCLA}
  \city{Los Angeles}
  \country{USA}}
\email{chenfanfu.jiang@gmail.com}

\author{Peter Yichen Chen}
\affiliation{%
  \institution{UBC}
  \city{Vancouver}
  \country{Canada}}
\email{pyc@csail.mit.edu}

\author{Yin Yang}
\affiliation{%
  \institution{University of Utah}
  \city{Salt Lake City}
  \country{USA}}
\email{yin.yang@utah.edu}


\begin{abstract}
Simulating large-scale articulated assemblies poses a significant challenge due to the numerical stiffness and geometric complexity of jointed structures. Conventional rigid body solvers struggle with the high nonlinearity induced by rotation parameterization. This difficulty becomes more pronounced for multiple two-way-coupled bodies. This paper introduces a novel framework that leverages the linear kinematic mapping of Affine Body Dynamics (ABD). As ABD targets near-rigid objects, the constitutive variations of different materials become negligible, which justifies a co-rotational approach to isolate geometric nonlinearities of the system. This insight enables the use of constant system matrices that can be pre-factorized throughout the simulation, even with fully implicit integration schemes. To manage the high DOF counts of large-scale systems, we map primal body coordinates onto a compact dual space defined by minimal joint degrees of freedom. By solving the resulting KKT systems, our method ensures exact constraint enforcement and physically accurate motion propagation. We provide a suite of specialized solvers tailored for diverse joint topologies, including chains, trees, closed loops, and irregular networks. Experimental results show that our approach achieves interactive rates for systems with hundreds of thousands of bodies on a single CPU core, while maintaining excellent stability at large time steps.
\end{abstract}

\ccsdesc[500]{Computing methodologies~Physical simulation}

\keywords{Rigid body dynamics, Multibody dynamics, Reduced model, Kinematic constraints}

\begin{teaserfigure}
\centering
  \includegraphics[width=\textwidth]{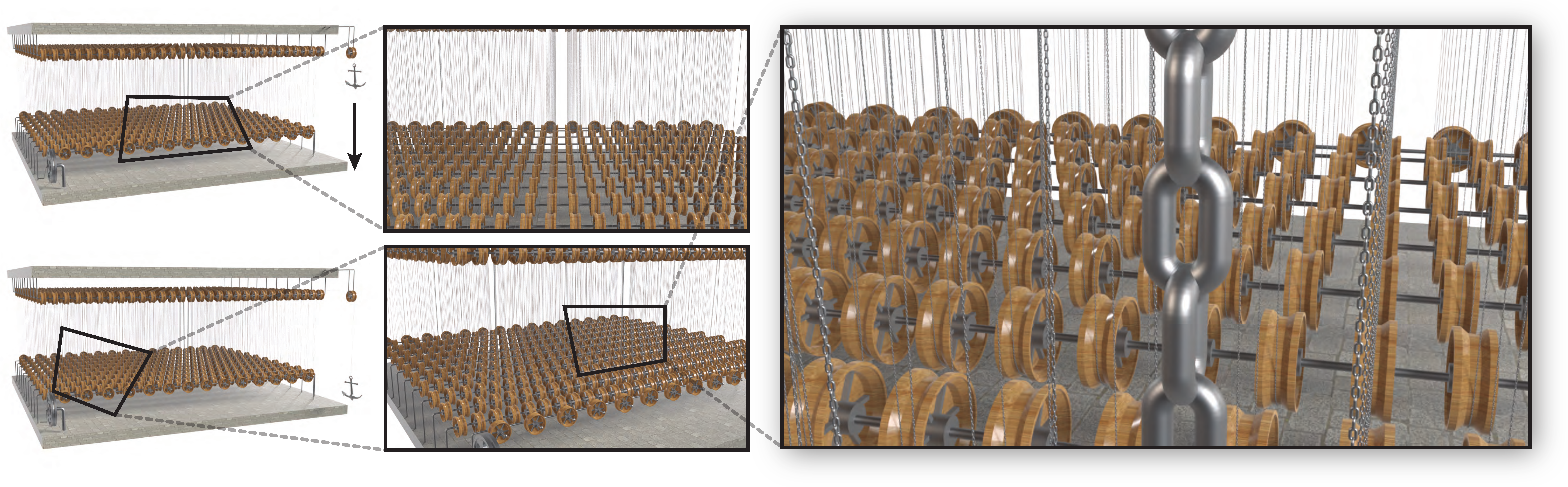}
\caption{\textbf{A huge pulley system.}
This paper proposes an affine body dynamics (ABD) based framework that integrates multiple rigid/near-rigid links via kinematic joints. Such a multi-ABD system is designed based on a smart co-rotated formulation of a single affine body. This strategy makes an ABD system matrix pre-factorized even with the implict integration. We then generalize all the commonly-seen joint constraints to the affine coordinate using either a linear or a nonlinear formulation. When joint constraints are exactly enforced with KKT, we can project any complex multibody system into the dual space with minimized freedoms. As most articulated systems are well-structured, we exploit topological connections among joints to fast solve complex systems. The teaser figure reports a representative example of a large-scale multibody system consisting of $1,076,748$ links. Our method robustly simulates this system at $h = 10^{-2}~s$ using only one iteration for each time step. The simulation takes $904~ms$ only with a single CPU thread.
}
\label{fig:huge_chain}
\Description{}
\end{teaserfigure}

\maketitle

\section{Introduction}\label{sec:intro}
Rigid bodies and articulated multibody systems are central to computer graphics, robotics, fabrication, and control. Multibody dynamics is formulated based on the representation of individual rigid bodies and the mathematical modeling of their joint constraints. In the traditional rigid body dynamics (RBD) framework~\cite{baraff1997introduction,erleben2005physics}, a rigid object is defined by a minimal set of six degrees of freedom (DOFs), representing its global translation and rotation. This formulation assumes that the body's local geometry remains undeformed under any external forces or joint interactions. The material is therefore considered infinitely stiff. A hidden drawback of RBD is that the relation between the spatial coordinates, i.e., $x$, $y$, $z$ coordinates, and RBD coordinates, i.e., a certain parameterization of the rotation and translation vector, is nonlinear. This nonlinearity may be of less importance for a single body. However, in the context of a multibody system, the time-varying constraint Jacobian leads to a time-varying system matrix if the joint constraints need to be exactly enforced, e.g., via KKT (Karush-Kuhn-Tucker) systems~\cite{boyd2004convex}. Existing methods often rely on explicit integration with highly conservative time steps or employ penalty methods to approximate constraints, while fully implicit RBD-multibody simulation is less common. As the system scales up, these methods fail to enforce joint constraints, leading to numerical instability and constraint drift.

Affine Body Dynamics~\cite{ABD} serves as a more versatile alternative. By expanding the kinematic space from 6 DOFs to 12 DOFs, ABD replaces nonlinear rotations with a linear mapping. This linearity ensures that the Jacobian remains constant w.r.t. the body's spatial coordinates, whereas the nonlinear rigidity is enabled with material stiffness. The resultant material nonlinearity requires the re-assembly and factorization of the system matrix at every iteration. As the kinematic space is expanded to 12 DOFs, ABD inevitably incurs higher computational costs than the RBD model.

We propose a scalable multibody simulation framework based on a dedicated numerical treatment of ABD. Our approach leverages a key observation: high material stiffness suppresses \emph{constitutive nonlinearity}, and the key challenge lies in the \emph{geometric nonlinearity}. Fortunately, it is known that difficulty can be resolved effectively through a co-rotational formulation~\cite{muller2004interactive}. By extracting the body's rotation, we land a constant system matrix that can be pre-factorized even using an implicit time integration scheme. This strategy serves as the core building block of our framework. 

We provide a detailed analysis of commonly used joint constraints under the affine coordinate, and detail their possible linear and/or nonlinear formulations. For multibody systems consisting of a large number of stiff components, rather than resorting to penalty-based constraint approximation, we exactly enforce joint constraints via the KKT form to ensure accuracy and stability of the simulation. Since each body's primal system is pre-factorized, projecting the global primal problem onto the dual space becomes highly efficient. By adopting a semi-linearization for each joint with minimal DOFs, we reach a dual problem whose dimension is significantly smaller than that of the original primal version. We propose efficient algorithms for different mechanical structures, including a block-tridiagonal solver, an extension of the Featherstone algorithm~\cite{featherstone2008rigid} for ABD-based multibody systems, a low-dimensional inverse approximation when kinetic loops exist, and a bidirectional Gauss-Seidel optimizer for dense joint networks.

We conduct extensive experiments to validate this new framework. Our method outperforms existing rigid body simulators and algorithms for both accuracy and efficiency. The proposed approach demonstrates excellent robustness, maintaining stability at a time step of $0.01~sec$ even for systems containing one million bodies. Because the joint constraints are strictly enforced via the KKT form, our simulator accurately propagates motion across the entire system. Even on a single-threaded CPU, we enable interactive simulation of highly complex multibody systems.  
\vspace{-10 pt}
\section{Related Work}

Articulated rigid-body simulation has a long history in graphics and robotics, with a central challenge being how to enforce joint constraints and contacts robustly at scale.
Classical rigid-body contact can be formulated as a Linear Complementarity Problem (LCP) dating back to early work on unilateral constraints and non-penetration \cite{Moreau1985,Baraff1989}, and many subsequent methods focus on improving robustness and efficiency of constraint solves and impact handling \cite{Stewart2000,Smith2012}.
For articulated systems in particular, prior work explored impulse-based formulations for contact and joints \cite{Mirtich1994,Weinstein2006}, adaptive or reduced representations for speed \cite{Redon2005}, and scalable treatments of highly constrained motion \cite{Sueda2011}.
Recent improvements include stabilization and geometric-acceleration ideas for constrained dynamics \cite{Tournier2015,Andrews2017}, as well as structured factorizations and Schur-complement techniques that exploit articulated sparsity \cite{Enzenhofer2019,Peiret2019}.
These approaches highlight the importance of exploiting kinematic structure when solving large constrained systems, particularly for chains, trees, and lightly loopy mechanisms.

A canonical example of exploiting kinematic structure is the articulated-body algorithm (ABA) introduced by Featherstone \cite{featherstone2008rigid}.
ABA achieves linear-time complexity for tree-structured articulated systems by recursively propagating articulated inertia and bias forces along the kinematic hierarchy.
Subsequent work extends this idea to implicit integration, loop handling, and constraint-consistent formulations, and ABA-style recursion remains a foundational building block in robotics and multibody dynamics.
Our solver draws conceptual inspiration from this line of work, but adapts the recursion and condensation strategy to affine generalized coordinates and KKT-based constraint projection.

Practical engines and toolkits such as MuJoCo, and other widely-used multibody platforms emphasize numerical robustness, efficient sparse solves, and engineered contact handling for application-driven workloads \cite{todorov2012mujoco}.
These implementations illustrate important trade-offs between representation choice, solver complexity, and runtime performance that guide design choices in both graphics and robotics-grade simulators.

Beyond rigid-link mechanisms, many graphics phenomena with long, filamentary structures (e.g., cables, hair, vegetation) are often modeled using continuum-inspired reduced formulations. Work on discrete elastic rods, strands, and vegetation emphasizes specialized reduced models and continuum mechanics to achieve efficiency for very long or highly elastic structures \cite{DER}.
These models focus on elastic deformation rather than exact articulated joints.
Our articulated ABD formulation complements this literature by targeting scenarios in which rigid-link connectivity, exact joint enforcement, and large-scale constraint propagation dominate the computational cost.

Efficient articulated solvers in maximal or hybrid coordinates have also been developed with near-linear scaling in the number of bodies for many practical scenes \cite{Wang2019RedMax}.
Related efforts in robotics further explore structure-aware and parallel variants of articulated-body algorithms to improve scalability and numerical robustness in deep kinematic trees.
Differentiable articulated simulators broaden these formulations to learning, system identification, and control, where stability and solver structure are critical \cite{Werling2021,Geilinger2020ADD,Macklin2019NonSmoothNewton}.
Complementary to these lines, quaternion-based constrained rigid-body formulations provide an alternative, implementation-friendly parameterization and constraint handling strategy \cite{VersatileQuaternion}.
Many production simulators adopt semi-implicit stepping with stabilized velocity constraints (e.g., Baumgarte-type stabilization), which can exhibit drift and requires careful tuning; implicit position-level formulations instead solve constraints to a prescribed tolerance \cite{Baumgarte1972,Erez2015}.

Beyond rigid-only pipelines, coupling rigid and deformable bodies introduces heterogeneous degrees of freedom and typically demands special-case treatments \cite{Shinar2008,Kaufman2008,JainLiu2011,BaiLiu2014}.
Unified constraint views such as position-based dynamics can couple rigid and deformable components with customized iterative solvers, but do not generally provide strong global guarantees or exact constraint satisfaction \cite{PBD,Deul2016,Muller2020}.
In contrast, optimization-based time integration has recently enabled stronger robustness properties for contact and constraints.
Incremental Potential Contact (IPC) formulates contact (and friction) via barrier energies inside a variational integrator, yielding intersection-free behavior under appropriate numerics \cite{IPC}, and has been extended to codimensional objects \cite{CIPC}.
For rigid bodies, intersection-free formulations directly in $\mathrm{SE}(3)$ have also been studied \cite{rigidIPC}, offering a complementary perspective on robust contact handling through geometric integration.

Geometric and variational integrators on Lie groups preserve structure such as momentum, symplectic form, and orthogonality, and provide desirable long-term behavior for articulated mechanisms \cite{Marsden_West_2001,Hairer2006}.
While attractive theoretically, these methods are typically formulated in minimal-coordinate $\mathrm{SE}(3)$ settings and introduce configuration-dependent Jacobians for constraint enforcement.
By contrast, our affine co-rotational approach trades a modest increase in degrees of freedom for constant kinematic mappings that interact more favorably with KKT-based condensation.

Our work is most closely related to affine-coordinate rigid modeling and optimization-based multibody frameworks.
Affine Body Dynamics (ABD) represents each rigid body with a $12$-DoF affine generalized coordinate and enforces rigidity via an orthogonality energy, while benefiting from Euclidean trajectories that simplify continuous collision detection and contact handling in collision-heavy scenes \cite{ABD}.
Compared to minimal-coordinate formulations on $\mathrm{SE}(3)$, affine coordinates trade a modest increase in per-body degrees of freedom for linear kinematic mappings and constant Jacobians, which can significantly simplify constraint assembly and projection.
Chen et al.\ propose a unified Newton-barrier multibody framework that combines optimization-based time integration, IPC-style contact handling, and a general treatment of articulation constraints, including an efficient change-of-variables approach for linear equality constraints \cite{UnifiedNewtonBarrier}.

Building on these foundations, our focus is on articulated rigid-body simulation with affine generalized coordinates while designing a solver that is simple to author, preserves key geometric invariants under large rotations, and exploits problem structure so that the dominant computational cost depends on constraint connectivity rather than the total state dimension.
In particular, we draw inspiration from articulated-body algorithms and Schur-complement-based elimination, but tailor them to the affine setting so that per-body systems can be pre-factorized and reused, enabling efficient dual-space solves for chains, trees, and more general joint graphs.

\vspace{-5 pt}
\section{Single-object Stiff Dynamics}
To make our exposition self-contained, we start with a brief review of RBD~\cite{baraff1997introduction} and ABD~\cite{ABD} for a single rigid or near-rigid object. Classic RBD defines the position of any material point as $\bm{x}_i = \bm{R}\bar{\bm{x}}_i + \bm{t}$ for a given rotation matrix $\bm{R}$ and translation $\bm{t}$. Here, $\bm{x}_i$ and $\bar{\bm{x}}_i$ denote the displaced and rest-shape positions of point $i$. In other words, the rigidity of the object is \emph{kinematically} prescribed.


\subsection{ABD Kinematics}
ABD allows the object to deform under a uniform linear transformation $\bm{A}(t)$ and a translation $\bm{t}(t)$ such that:
\begin{equation}\label{eq:abd}
   \bm{x}_i = \bm{A}(t)\bar{\bm{x}}_i + \bm{t}(t).
\end{equation}
The rigidity of the object is approached with an elasticity potential of:
\begin{equation}\label{eq:Eabd}
    E_A = \int_{\Omega} \Psi d \Omega.
\end{equation}
$\Psi$ is the energy density function, and many known hyperelastic models can be used as long as they are rotation-invariant. For instance, the vanilla ABD~\cite{ABD} used a simple polynomial energy of:
\begin{equation}\label{eq:EA}
    \Psi = k_A  \left\| \bm{A} \bm{A}^\top - \bm{I}_3 \right\|^2_F,
\end{equation}
where $\| \cdot\|_F^2$ is the Frobenius norm; $\bm{I}_3$ is a $3 \times 3$ identity matrix; and $k_A$ is the affine stiffness of the body. 

The relaxation from $\bm{R} \in SO(3)$ to $\bm{A}$ offers both advantages and disadvantages compared with RBD. On the positive side, the affine DOFs can be used to store elastic energies, which facilitates energy-based collision and contact processing such as Incremental Potential Contact (IPC)~\cite{IPC}. The nodal positions $\bm{x}$ a.k.a. the \emph{spatial coordinates} are linearly related to the affine DOFs such that:
\begin{equation}\label{eq:abd_dof}
\bm{x}_i = \bm{J}_i\bm{q} = [\bar{\bm{x}}_i^\top \otimes \bm{I}_3, \bm{I}_3] \bm{q} \Leftrightarrow \bm{x} = \underbrace{[\bm{J}_1^\top,...,\bm{J}_N^\top]^\top}_{\bm{J}} \bm{q}
\end{equation}
where the generalized affine coordinate $\bm{q} = [\operatorname{vec}^\top(\bm{A}), \bm{t}^\top]^\top \in \mathbb{R}^{12}$ is constructed by stacking the three column vectors of $\bm{A}$ and the translation vector $\bm{t}$. $\bm{J}$ is the coordinate Jacobian and is constant when the rest-shape model is given.

At the other end of the spectrum, ABD needs six more DOFs than RBD, which doubles the size of the system matrix. At each time step, we need to assemble and factorize a $12 \times 12$ system matrix. This is notably slower than an RBD system. Therefore, it is often believed that the multi-ABD framework is less scalable than the multi-RBD framework, especially when there exists a large number of interconnected stiff objects/links.


\subsection{ABD Dynamics}
Assuming the affine body is discretized with a tetrahedral mesh of $N$ nodes, we can formulate its equilibrium at each time step as a variational optimization of:
\begin{equation}\label{eq:variation_energy}
    \bm{x}^{n + 1} = \arg\min_{\bm{x}} E(\bm{x}), \quad E(\bm{x}) = E_I (\bm{x}) + \Psi(\bm{x}),
\end{equation}
where the subscript $n+1$ indicates the time step index. $E_I$ is the inertia potential defined as:
\begin{equation}
    E_I(\bm{x}) = \frac{1}{2h^2} (\bm{x} - \hat{\bm{x}})^\top \bm{M} (\bm{x} - \hat{\bm{x}}). \nonumber
\end{equation}
$h$ is the time step size. $\bm{M}$ is the fullspace mass matrix. $\hat{\bm{x}}$ is a known vector depending on the kinematic state of the system from the previous time step. For instance if the implicit Euler is used, $\hat{\bm{x}} = \bm{x}^n + h\dot{\bm{x}}^n + h^2 \bm{M}^{-1}\bm{f}_{ext}$ with $\bm{f}_{ext}$ being the external force.

We use Newton's method to solve the optimization of Eq.~\eqref{eq:variation_energy}, leading to a linear system of:
\begin{equation}\label{eq:newton}
    \bm{H}\delta \bm{x} = -\bm{g},
\end{equation}
where 
\begin{equation}
    \bm{H} = \frac{\partial^2 E(\bm{x})}{\partial^2 \bm{x}} = \frac{1}{h^2}\bm{M} + \bm{K}(\bm{x}), \quad \bm{g} = \frac{\partial E(\bm{x})}{\partial \bm{x}} \nonumber
\end{equation} 
are the Hessian and gradient of the total variational energy. $\bm{K}(\bm{x}) = \partial^2 \Psi / \partial \bm{x}^2$ is a.k.a. the tangent stiffness matrix. 

We can then project Eq.~\eqref{eq:newton} into the column space of $\bm{J}$:
\begin{equation}\label{eq:newton_abd}
     \bm{H}_A \delta \bm{q} = - \bm{J}^\top \bm{g}, \quad \bm{H}_A = \frac{1}{h^2}\bm{M}_A + \bm{K}_A,
\end{equation}
where $\bm{M}_A = \bm{J}^\top \bm{M} \bm{J}$ is the generalized mass matrix. Unlike its counterpart in RBD, $\bm{M}_A$ is constant and is pre-computed. The generalized stiffness matrix $\bm{K}_A = \bm{J}^\top \bm{K}(\bm{x}) \bm{J}$ represents the main computation hurdle. Since \emph{any} rotation-invariant energy is nonlinear w.r.t. $\bm{x}$ (and thus w.r.t. $\bm{q}$), we need to re-assemble $\bm{K}$ at each Newton iteration, and project the resulting $\bm{K}$ to the affine subspace. 

\subsection{A Co-rotated Formula}
Eq.~\eqref{eq:newton_abd} solves a $12 \times 12$ linear system at each Newton step. Both its assembly and solve are considerably more expensive than its RBD counterpart. We show that it is possible to completely resolve the limitation and have ABD reclaim a performance lead over RBD via a co-rotated formulation.

Both RBD and ABD are nonlinear. Nevertheless, their sources of the nonlinearity are drastically different. Specifically, the nonlinearity of RBD lies in the map between the rigid coordinate and nodal positions. In contrast, the nonlinearity of ABD stems from the geometric nonlinearity of the stiff material i.e., the requirement of being rotation-invariant. 
In a multibody system, we focus on highly stiff objects, meaning the actual deformation of an affine body is small or nearly vanished. The small deformation prior plays a critical role: it suggests the strain-stress relations among different energy models are indistinguishable in the \emph{material space}, which is nearly identical to the one at the rest shape. Mathematically, it means:
\begin{multline}\label{eq:reduced_K}
    \bm{K}(\bm{x}) \delta \bm{x} \approx \operatorname{diag}_{N}(\bm{R}) \bar{\bm{K}} \big(\operatorname{diag}_{N}(\bm{R}^\top) \delta \bm{x}\big) \\
    =  \underbrace{\big( \operatorname{diag}_{N}(\bm{R}) \bar{\bm{K}} \operatorname{diag}_{N}(\bm{R}^\top) \big)}_{\tilde{\bm{K}}} \delta \bm{x},
\end{multline}
where $\operatorname{diag}_{N}(\bm{R}) \in \mathbb{R}^{3N \times 3N}$ is a block-diagonal matrix of $N$ copies of $\bm{R}(\bm{q}) \in \mathbb{R}^{3 \times 3}$. $\bm{R}(\bm{q})$ is the rotation matrix relating the object's rest-shape orientation and current orientation. It can be conveniently computed by applying the polar decomposition based on the current affine coordinate $\bm{q}$:
\begin{equation}\label{eq:polar}
   \bm{R}(\bm{q}) = \bm{A}(\bm{A}^\top \bm{A})^{-\frac{1}{2}}, \quad \bm{A}(\bm{q}) = \operatorname{vec}^{-1}\big([\bm{I}_9,  \bm{0}_{9 \times 3}] \bm{q}\big).
\end{equation}
$\tilde{\bm{K}}$ is a high-quality approximation of the ground-truth $\bm{K}(\bm{x})$. The approximation error in Eq.~\eqref{eq:reduced_K} vanishes as $\bm{A} \rightarrow \bm{R}$. Under the close-to-rigid assumption, \emph{this error is virtually a numeric zero in practice}.

For a whatever rotation $\bm{R}$, it is easy to see $\bm{R} \bm{x}_i  = (\bm{RA}) \bm{\bar{x}}_i + \bm{R} \bm{t} $ by left-multiplying $\bm{R}$ at both sides of Eq.~\eqref{eq:abd}. This leads to:
\begin{equation}
    \bm{x} = \bm{J} \bm{q} \Rightarrow \operatorname{diag}_{N}(\bm{R}) \bm{x} = \operatorname{diag}_N (\bm{R}) \bm{J} \bm{q} = \bm{J} \operatorname{diag}_4(\bm{R}) \bm{q} \nonumber,
\end{equation}
which holds for any $\bm{q}$, and we therefore have:
\begin{equation}\label{eq:abd_r}
    \operatorname{diag}_N (\bm{R}) \bm{J} = \bm{J} \operatorname{diag}_4 (\bm{R}).
\end{equation}
Eq.~\eqref{eq:abd_r} reveals an important feature of ABD: the spatial coordinate $\bm{x}$ and generalized coordinate $\bm{q}$ co-rotate i.e., with $\operatorname{diag}_N (\bm{R})$ and $\operatorname{diag}_4 (\bm{R})$ under the same global rotation. 

Bearing this property in mind, the reduced stiffness matrix can be simplified as:
\begin{multline}
\bm{K}_A = \bm{J}^\top \tilde{\bm{K}} \bm{J} = \bm{J}^\top \operatorname{diag}_{N}(\bm{R}) \bar{\bm{K}} \operatorname{diag}_{N}(\bm{R}^\top) \bm{J} \\ 
    = \operatorname{diag}_4(\bm{R}) \underbrace{\bm{J}^\top \bar{\bm{K}} \bm{J}}_{\bar{\bm{K}}_A} \operatorname{diag}_4(\bm{R}^\top) . 
\end{multline}
$\bar{\bm{K}}_A$ is the rest-shape generalized stiffness matrix, and it can be pre-computed. 

Similarly, the generalized mass is also rotation-invariant since:
\begin{multline}
    \bm{M}_A = \bm{J}^\top \bm{M} \bm{J} = \operatorname{diag}_4 (\bm{R}) \bm{J}^\top \operatorname{diag}_N (\bm{R}^\top) \bm{M} \operatorname{diag}_N (\bm{R})  \bm{J} \operatorname{diag}_4 (\bm{R}^\top)\\
    = \operatorname{diag}_4 (\bm{R}) \bm{M}_A \operatorname{diag}_4 (\bm{R}^\top).
\end{multline}
Eq.~\eqref{eq:newton_abd} can be re-written as:
\begin{equation}\label{eq:reduced_newton}
     \operatorname{diag}_4 (\bm{R})\left(\frac{1}{h^2}\bm{M}_A + \bar{\bm{K}}_A\right)\operatorname{diag}_4 (\bm{R}^\top) \delta \bm{q} = - \bm{J}^\top \bm{g}.
\end{equation}

The r.h.s. of Eq.~\eqref{eq:reduced_newton} integrates the elasticity gradient over each element $e$ on the mesh to the affine coordinate such that:
\begin{multline}
    -\bm{J}^\top \bm{g} = \sum_e \bm{J}_e^\top\frac{\partial \int_{\Omega_e} \Psi d {\Omega_e}}{\partial \bm{x}_e} = \sum_e V_e \bm{J}_e^\top \left(\frac{\partial \Psi}{\partial \bm{F}_e} :\frac{\partial \bm{F}_e}{\partial \bm{x}_e}\right)\\
    = \left(\sum_e V_e \bm{J}_e \right)^\top \left(\frac{\partial \Psi}{\partial \bm{A}} :\frac{\partial \bm{F}_e}{\partial \bm{x}_e}\right) = \bar{\bm{J}}^\top \bm{f} =  \bm{f}_A,
\end{multline}
where $\bm{F}_e \in \mathbb{R}^{3 \times 3}$ is the deformation gradient of the element. $\partial \bm{F}_e/\partial \bm{x}_e$ is a constant and sparse 3rd tensor for the linear tetrahedron (e.g., see~\cite{kim2020dynamic}). For an ABD mesh, we have $\bm{F}_e = \bm{A}$ making $\partial \Psi/ \partial \bm{F}_e = \partial \Psi/ \partial \bm{A}$ constant across all the elements. As a result, the double contraction can be moved out of the summation sign. $\bar{\bm{J}} \in \mathbb{R}^{12 \times 12}$ is volume-weighted Jacobian, which can be pre-computed. It converts the aggregated spatial force $\bm{f} \in \mathbb{R}^{12}$ to the affine coordinate $\bm{f}_A$. Therefore, the complexity of computing the r.h.s. of Eq.~\eqref{eq:reduced_newton} is independent of the mesh resolution. 

Because of the small deformation prior, we opt for the linear material model for $\Psi$. The affine energy used in~\cite{ABD} i.e., Eq.~\eqref{eq:Eabd} can also be used, but it fails to incorporate both Lam\'e parameters ($\mu$, $\lambda$). The simplification of linear elasticity gives:
\begin{equation}
    \frac{\partial \Psi}{\partial \bm{A}} = \mu(\bm{A} + \bm{A}^\top - 2\bm{I}_3) + \lambda \operatorname{tr}(\bm{A} - \bm{I}_3) \bm{I}_3.
\end{equation}
Linear elasticity offers the first-order approximation of most rotation-invariant hyperelastic models around the rest shape. Because each Newton iteration also linearizes the equation of motion, \emph{the error induced by material simplification is hidden by the error induced by the Newton linearization}. As a result, the convergence of Newton's method is never impacted by material simplification.

\subsection{Single-object ABD}
The final system for a single-body ABD now becomes:
\begin{equation}\label{eq:abd_final}
    \bar{\bm{H}}_A\operatorname{diag}_4 (\bm{R}^\top) \delta \bm{q} =  \operatorname{diag}_4 (\bm{R}^\top) \bm{f}_A, \quad \bar{\bm{H}}_A = \frac{1}{h^2}\bm{M}_A + \bar{\bm{K}}_A.
\end{equation}
Eq.~\eqref{eq:abd_final} features a constant $\bar{\bm{H}}_A$, which is pre-factorized. In other words, we avoid any matrix factorization, even with an implicit time integration. 

As the system matrix is pre-factorized, the polar decomposition that extracts the rotation component from the affine coordinate $\bm{q}$ i.e., Eq.~\eqref{eq:polar} becomes a major computational hurdle. We want to remind that simulating ABD using Eq.~\eqref{eq:abd_final} is already highly efficient and much faster than implicit RBD. Nevertheless, we can further push the performance by skipping Eq.~\eqref{eq:polar} and replacing it with a simple normalization operation. As the object is highly stiff, $\bm{A} \approx \bm{R}$ is already close to a rotation matrix, when $\bm{R}\bm{a}$ rotates a vector $\bm{a}$, instead of computing $\bm{R}$ out of $\bm{A}(\bm{q})$, we only enforce $\|\bm{A} \bm{a}\| = \| \bm{a}\|$ to make sure $\bm{A}$ preserves the length of $\bm{a}$ such that:
\begin{equation}\label{eq:len_preserving}
    \bm{Ra} \approx \frac{\| \bm{a} \|}{\|\bm{Aa} \|}\bm{A}\bm{a} .
\end{equation}
By skipping the polar decomposition with Eq.~\eqref{eq:len_preserving}, the ABD simulation procedure only involves lightweight level 1 and 2 BLAS operations, making implicit ABD outperform both implicit and explicit RBD. The pseudo code for one Newton iteration of a single object ABD is summarized in Alg.~\ref{alg:abd}.
\begin{algorithm}
\DontPrintSemicolon
\KwIn{$\bm{M}_A$, $\bar{\bm{K}}_A$, $\bm{q} = [\bm{q}_1^\top, \bm{q}_2^\top, \bm{q}_3^\top, \bm{q}_4^\top]^\top$, $\bar{\bm{J}}$, $\bm{f}$, $h$}
\KwOut{$\delta \bm{q}$}
\BlankLine
$\bm{H}_A \leftarrow \frac{1}{h^2}\bm{M}_A + \bar{\bm{K}}_A$\\
$\bm{f}_A = [\bm{f}_{A, 1}^\top, \bm{f}_{A, 2}^\top, \bm{f}_{A, 3}^\top, \bm{f}_{A, 4}^\top]^\top \leftarrow \bar{\bm{J}}^\top \bm{f}$\\
$\bm{A} \leftarrow [\bm{q}_1, \bm{q}_2, \bm{q}_3]$ \\
\For{$k = 1$  \KwTo $4$}
{
$l^2_{k} \leftarrow \bm{f}_{A, k} \cdot \bm{f}_{A, k}$\\
$\bm{f}_{A, k} \leftarrow \bm{A}^\top \bm{f}_{A, k}$ \\
$\bm{f}_{A, k} \leftarrow \sqrt{\frac{l_k^2}{\bm{f}_{A, k} \cdot \bm{f}_{A, k}}} \cdot \bm{f}_{A, k} $
}
Solve $\delta \bm{p} = [\delta \bm{p}_1^\top, \delta \bm{p}_2^\top, \delta \bm{p}_3^\top, \delta \bm{p}_4^\top]^\top$ via $\bm{H}_A \delta \bm{p} =  \bm{f}_A$ \\
\For{$k = 1$  \KwTo $4$}
{
$l^2_{k} \leftarrow \delta \bm{p}_k \cdot \delta \bm{p}_k$\\
$\delta \bm{p}_k \leftarrow \bm{A} \delta \bm{p}_k$ \\
$\delta \bm{q}_k \leftarrow \sqrt{\frac{l_k^2}{\delta \bm{p}_k \cdot \delta \bm{p}_k}} \cdot \delta \bm{p}_k $
}
\caption{One Newton-step ABD integration}\label{alg:abd}
\end{algorithm}

\begin{figure}
\centering
\includegraphics[width=\linewidth]{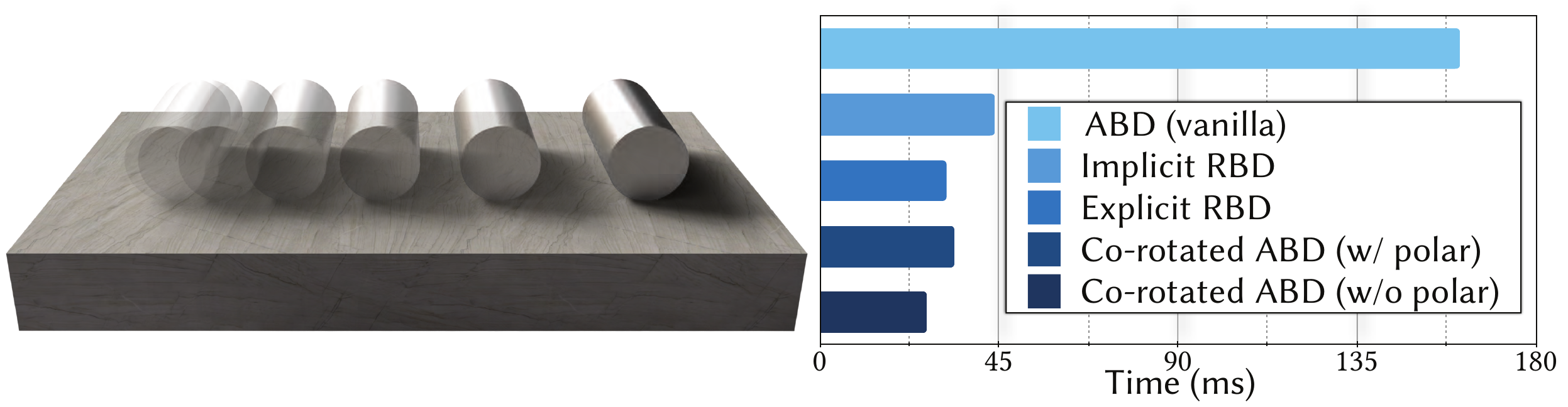}
\caption{\textbf{Performance comparison ABD vs. RBD.}~~We simulate a rolling cylinder for $10$K steps and compare the total simulation using several variations of RBD and ABD. As visualized on the right, the vanilla implicit ABD is the least efficient ($161~ms$) as it needs to assemble and solve a $12 \times 12$ matrix for each iteration. It is much slower than implicit RBD ($44~ms$) and explicit RBD ($32~ms$). With our co-rotated formulation, the system matrix for implicit ABD is pre-factorized, which significantly reduces the computation time from $161~ms$ to $34~ms$. By skipping polar decomposition, the performance of (implicit) ABD reaches $27~ms$, which is over $20\%$ faster than explicit RBD. The experiment is tested on an i7 CPU with single thread.}\label{fig:performance}
\Description{}
\end{figure}

A simple benchmark test is reported in Fig.~\ref{fig:performance}, where we simulate a rigid rolling cylinder for $10K$ step (with $h = 0.01~sec$) with ABD and RBD. The vanilla implicit ABD i.e., the original algorithm proposed in~\cite{ABD} uses $161~ms$, while implicit and explicit RBD need $44~ms$ and $32~ms$, respectively. Our co-rotated ABD (with polar decomposition) takes $34~ms$, which is nearly as fast as explicit RBD already. Skipping polar decomposition further shortens the total computation time to $27~ms$.

\vspace{-5 pt}
\section{Multi-object ABD}\label{sec:multiabd}
The co-rotated formulation eliminates the major disadvantage of ABD. Based on this, we show how to incorporate different kinematic joints to assemble a multi-ABD system while maintaining computational efficiency. Without loss of generality, we consider basic cases where two affine bodies are constrained with a bilateral joint. 
\vspace{-20 pt}
\setlength{\columnsep}{5 pt}
\begin{wrapfigure}{l}{0.5\linewidth}
\includegraphics[width=\linewidth]{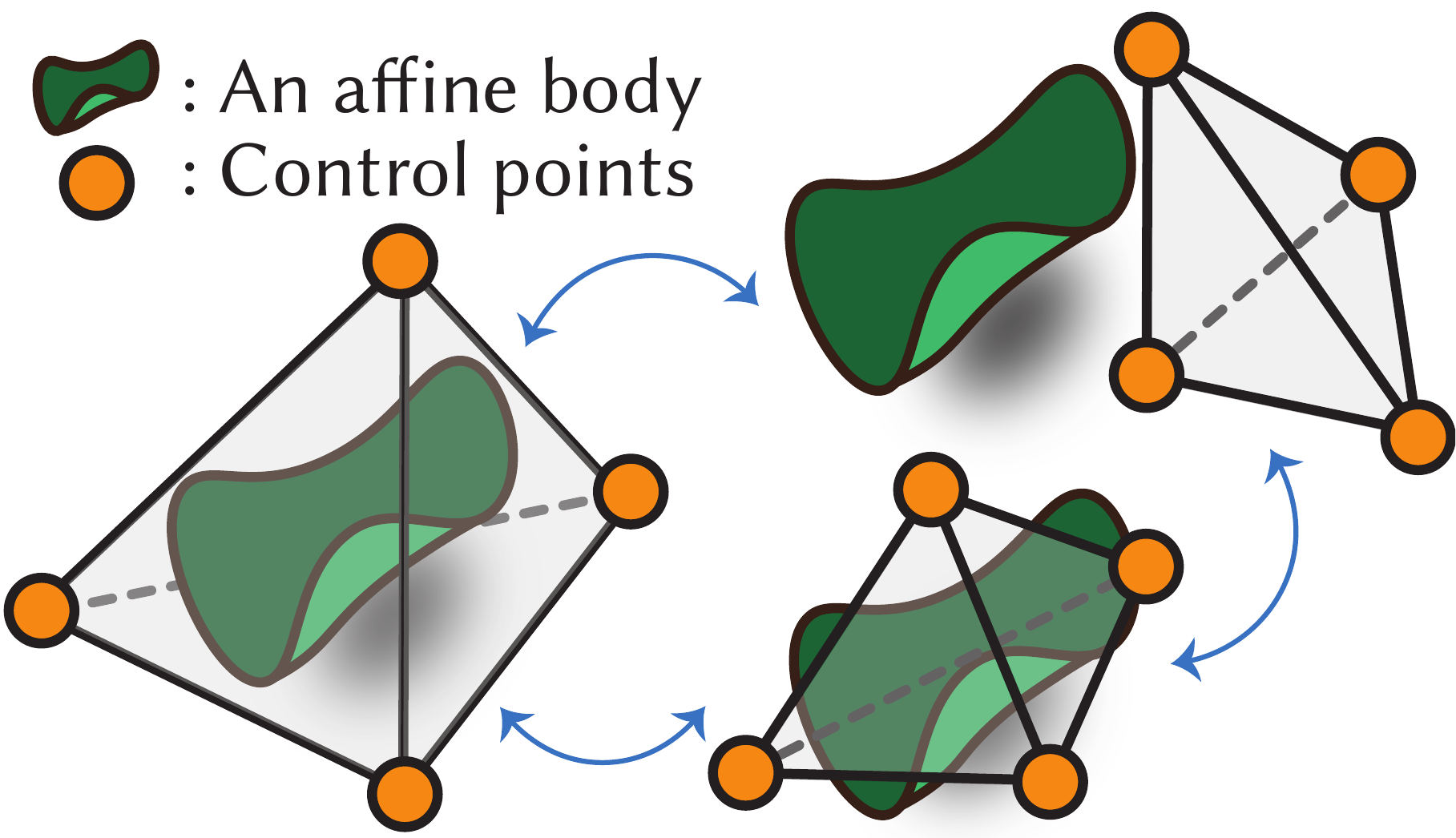}
\caption{\textbf{Control points and tetrahedron.} We can map the generalized coordinate of an affine body to any non-degenerate tetrahedron or the control tetrahedron. Its four corners are named control points of the affine body. Concatenating their positions gives the CP coordinate.} \label{fig:cp}
\end{wrapfigure}
\vspace{-22 pt}
\subsection{Control points and control tetrahedron of ABD}
We re-parameterize the affine coordinate $\bm{q}$ as a collection of four \emph{control points} (CPs), which forms a bijective map between $\bm{q}$ and CPs' spatial coordinates.

\begin{figure*}
    \centering
    \includegraphics[width=\linewidth]{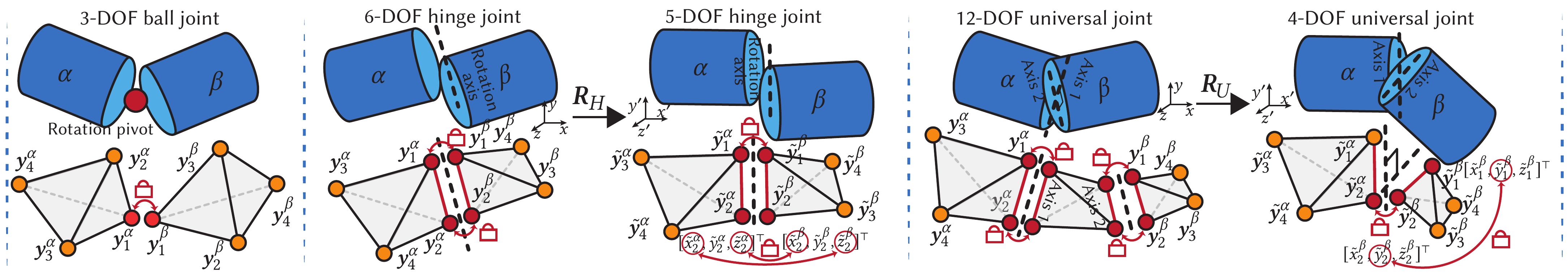}
    \caption{\textbf{Rotational joints.} The affine coordinate linearizes rotational constraints between affine bodies. By pre-parameterizing $\bm{q}$ to the CP coordinate $\bm{y}$, a ball joint becomes a point-point constraint; a hinge joint can be modeled as a 6-DOF edge-edge constraint; and a universal joint is the superposition of two hinge joints (with a virtual intermediate body). If we transfer the interconnected bodies into the local frame, the hinge joint can be enforced with 5 DOFs, and the universal joint only consumes 5 DOFs i.e., with the minimum DOFs needed.}
    \label{fig:joint}
\end{figure*}
Let $\bar{\bm{y}}_1$, $\bar{\bm{y}}_2$, $\bar{\bm{y}}_3$, $\bar{\bm{y}}_4$ be the spatial coordinates of four nodes of a tetrahedron corresponding to the rest-shape state of an affine body, and we name this tetrahedron the control tetrahedron (CT). When an affine deformation $\bm{q} = [\operatorname{vec}^\top(\bm{A}), \bm{t}^\top]^\top$ occurs, the deformed positions of CPs are $\bm{y}_1$, $\bm{y}_2$, $\bm{y}_3$, and $\bm{y}_4$. They are concatenated as $\bm{y} = [\bm{y}^\top_1, \bm{y}^\top_2, \bm{y}^\top_3, \bm{y}^\top_4]^\top \in \mathbb{R}^{12}$, which serves as another set of generalized coordinate or the \emph{CP coordinate} of the affine body. It is easy to verify that:
\begin{equation}\label{eq:cp_coordinate}
\bm{y} = \left[ \bm{Y}^\top \otimes \bm{I}_3,\;\bm{1}_{4\times1} \otimes \bm{I}_3\right]\bm{q} = \bm{T} \bm{q},\quad \bm{Y} = [\bar{\bm{y}}_1, \bar{\bm{y}}_2, \bar{\bm{y}}_3, \bar{\bm{y}}_4].
\end{equation}
As long as the CT is non-degenerate, we have $\bm{q} = \bm{T}^{-1} \bm{y}$,
meaning the Jacobian matrix $\frac{\partial \bm{q}}{\partial \bm{y}} = \bm{T}^{-1}$ is constant.

The choice of CT is flexible. As shown in Fig.~\ref{fig:joint}, the affine body can be fully enclosed by the CT, or partially overlap with it, or is completely separate from the CT. We then define joint constraints via the CP coordinate (instead of $\bm{q}$).


\subsection{Ball joint}
A ball joint eliminates relative translational motion between two stiff bodies $\alpha$ and $\beta$, while permitting their rotations. Let $\bm{y}_k^\alpha$ and $\bm{y}_k^\beta$ be their CPs with $k = 1, 2, 3, 4$. We enforce the ball constraint by setting one of the CPs of each body, say $\bm{y}_1^\alpha$ and $\bm{y}_1^\beta$, 
to be at the prescribed location of the ball joint. It becomes a positional constraint using with CP coordinate:
\begin{equation}\label{eq:ball}
    \bm{S}^\alpha_B \bm{y}^\alpha - \bm{S}_B^\beta \bm{y}^\beta = \bm{0} \Leftrightarrow \bm{S}_B^\alpha \bm{T}^\alpha \bm{q}^\alpha - \bm{S}_B^\beta \bm{T}^\beta \bm{q}^\beta = \bm{0},
\end{equation}
where we use the superscript to denote the body index in the system. $\bm{S}_B \in \mathbb{R}^{3 \times 12}$ is a selection matrix picking the pertaining DOFs out of the CP coordinate $\bm{y}$. A ball joint is a 3-DOF constraint. As a result, two ball-joint-connected affine bodies possess $12 + 12 - 3 = 21$ DOFs in total.

\subsection{Hinge joint}
A hinge joint or a revolute joint permits one rotational DOF i.e., the rotation about a specified axis, while all translational motion and all other rotations are constrained.

The relaxation to affine DOFs makes the hinge joint formulation handy. After mapping $\bm{q}$ to $\bm{y}$, one can enforce a hinge joint by requiring two edges on the CTs of two stiff bodies to align with the prescribed rotation axis, with their ends overlapped. In other words, the hinge joint becomes the superposition of two ball joints, and the corresponding affine hinge is a 6-DOF constraint (as shown in Fig.~\ref{fig:joint}). This strategy was used in previous ABD-based frameworks such as~\cite{UnifiedNewtonBarrier} as it is particularly implementation-friendly.

Alternatively, we propose a more compact encoding of a hinge joint, keeping its dimensionality the same as its RBD counterpart. As illustrated in Fig.~\ref{fig:joint}, we compute a hinge rotation $\bm{R}_H$, which aligns the prescribed rotation axis upright along the local $y$ axis:
\begin{equation}
 \bm{R}_H = \bm{I}_3 + [\bm{v}]_\times + \frac{[\bm{v}]_\times^2}{1 + a_y}, 
\end{equation}
where $\bm{a} =[a_x, a_y, a_z]^\top$ is the unit rotation axis of the hinge. $\bm{v} = \bm{a} \times \bm{e}_y$, where $\bm{e}_y = [0, 1, 0]^\top$.

\setlength{\columnsep}{5 pt}
\begin{wrapfigure}{l}{0.12\linewidth}
\includegraphics[width=\linewidth]{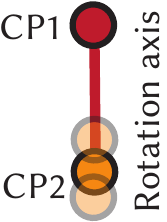}
\end{wrapfigure}
$\bm{R}_H$ allows us to convert $\bm{y}$ to the local frame: $\tilde{\bm{y}} = \operatorname{diag}_4(\bm{R}_H) \bm{y}$. With the local CP coordinate, the hinge constraint includes a ball constraint, fixing one CP from each body on the rotation axis. The other two constraints apply to the other CP from each body, which is also on the axis, but only at their local $x$ and $z$ coordinates. In other words, it restricts the movement of the CP to be only along the rotation axis (i.e., see the inset). Unlike in RBD-based multibody formulation, the distance between two CPs is not explicitly preserved. Instead, the distance change will be strongly penalized by the material stiffness. Therefore, the hinge constraint can be written as a 5-dimensional system:
\begin{equation}\label{eq:hinge}
    \bm{S}_H^\alpha \operatorname{diag}_4(\bm{R}_H) \bm{T}^\alpha \bm{q}^\alpha - \bm{S}_H^\beta \operatorname{diag}_4(\bm{R}_H) \bm{T}^\beta \bm{q}^\beta = \bm{0},
\end{equation}
where $\bm{S}_H \in \mathbb{R}^{5 \times 12}$ is the corresponding selection matrix.

\subsection{Universal joint}
A universal joint constrains all translational motion and one rotational motion, while permitting rotation about two orthogonal axes. A convenient implementation of the universal joint may be treating it as the composition of two perpendicular hinge joints. As shown in Fig.~\ref{fig:joint}, we can insert a virtual affine body whose CT has two orthogonal edges aligned with the two rotation axes of the universal joint. The universal constraint can then be enforced by adding two hinges at those two edges. This strategy needs 12 more primal DOFs for the added stiff body and consumes 12 DOFs. The total DOFs of the resulting system are $12 + 12 + 12 - 12 = 24$.

A compact formulation is also possible for the universal joint, which only needs a minimum set of $4$ DOFs. Let $\bm{a}_1 \perp \bm{a}_2$ be a pair of orthogonal axes of the joint. We can build a rotation matrix $\bm{R}_U$ via:
\begin{equation}
    \bm{R}_U = [\bm{a}_1 \times \bm{a}_2, \bm{a}_1, \bm{a}_2]^\top,
\end{equation}
such that $\bm{R}_U\bm{a}_1$ and $\bm{R}_U\bm{a}_2$ align with the local $y$ axis and $z$ axis respectively. Within the local frame, the universal joint can be realized as the composition of a ball joint and another equality constraint to make sure $\bm{a}_2$ can only move on the tangent plane of $\bm{a}_1$, whereas the length of $\bm{a}_2$ is maintained via the material stiffness. 

\begin{wrapfigure}{l}{0.26\linewidth}
\includegraphics[width=\linewidth]{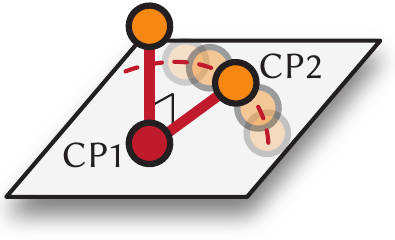}
\end{wrapfigure}
Take the setup in Fig.~\ref{fig:joint} as an example. Since $\bm{R}_U$ is built to rotate $\bm{a}_1$ and $\bm{a}_2$ to be parallel to local $y$ and $z$ axes, the CP-CP constraint of $\tilde{\bm{y}}^\alpha_2 = \tilde{\bm{y}}^\beta_2$ removes the relative translational movement between $\alpha$ and $\beta$. The additional equality constraint keeps the local $y$ coordinates of the first and the second CPs of $\beta$ identical i.e., $\tilde{y}_1^\beta = \tilde{y}_2^\beta$. This constraint restricts $\tilde{\bm{y}}^\beta_2$ to always reside in the local $\tilde{x}\tilde{z}$ plane and hence to be orthogonal to $\bm{a}_1$ (see the inset).  

The formulation of a universal joint is then:
\begin{equation}\label{eq:universal}
    \bm{S}_U^\alpha \operatorname{diag}_4(\bm{R}_U) \bm{T}^\alpha \bm{q}^\alpha - \bm{S}_U^\beta \operatorname{diag}_4(\bm{R}_U) \bm{T}^\beta \bm{q}^\beta = \bm{0}.
\end{equation}
Here, $\bm{S}_U \in \mathbb{R}^{4 \times 12}$ only picks four pertaining local CP coordinates.

\subsection{Prismatic joint}
When two bodies are connected by a prismatic joint, only relative translation along a prescribed axis is allowed. Similar to the hinge joint, we can find a rotation $\bm{R}_P$ such that the translational direction $\bm{a} = [a_x, a_y, a_z]^\top$ is parallel to the local $y$ axis:
\begin{equation}
\bm{R}_P = \bm{I}_3 + [\bm{v}]_\times + \frac{[\bm{v}]_\times^2}{1 + a_y}.
\end{equation}

\begin{wrapfigure}{r}{0.36\linewidth}
\vspace{-10 pt}
\includegraphics[width=\linewidth]{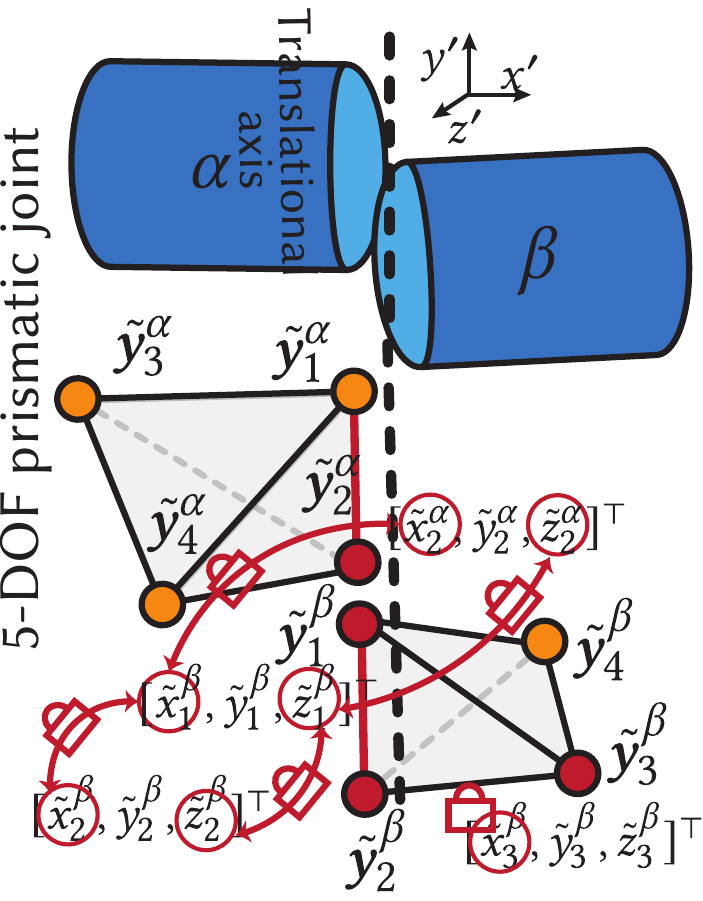}
\caption{\textbf{Prismatic joint.} A prismatic joint filters out-of-axis movements. The last constraint keeps $\beta$ non-rotatable. }
\label{fig:prismatic}
\end{wrapfigure}
A prismatic joint in the framework of ABD constrains a different set of local CT coordinates than a hinge joint. As shown in Fig.~\ref{fig:prismatic}, we initialize CTs so that two edges align with the translational direction $\bm{a}$. After $\bm{R}_P$ is applied, we enforce the local $\tilde{x}$ and $\tilde{z}$ coordinates from $\tilde{\bm{y}}^\alpha_2$, $\tilde{\bm{y}}^\beta_1$, and $\tilde{\bm{y}}^\beta_2$ be the same. This leads to four constraint equations in $\tilde{\bm{y}}$ coordinate: $\tilde{x}_2^\alpha - \tilde{x}_1^\beta = 0$, $\tilde{z}_2^\alpha - \tilde{z}_1^\beta = 0$, $\tilde{x}_1^\beta - \tilde{x}_2^\beta = 0$, and $\tilde{z}_1^\beta - \tilde{z}_2^\beta = 0$. The last constraint equation requires either $\tilde{\bm{y}}_4^\beta$ or $\tilde{\bm{y}}_3^\beta$ to stay with the current plane perpendicular to $\bm{a}$. To this end, we can simply fix $\tilde{x}$ or $\tilde{z}$ coordinate of any of them. In the figure, we set $\tilde{x}_3^\beta = 0$. Therefore, the constraint formulation of the prismatic joint is:
\begin{equation}\label{eq:prismatic}
    \bm{S}_P^\alpha \operatorname{diag}_4(\bm{R}_P) \bm{T}^\alpha \bm{q}^\alpha - \bm{S}_P^\beta \operatorname{diag}_4(\bm{R}_P) \bm{T}^\beta \bm{q}^\beta = \bm{0}.
\end{equation}
The selection matrices  $\bm{S}_P \in \mathbb{R}^{5 \times 12}$ has different structures on $\alpha$ and $\beta$.

\subsection{Linear or nonlinear}
In the classic RBD-based multibody system, the constraint equations for all types of kinematic joints in RBD are nonlinear. We say a constraint type is linear or nonlinear, meaning its gradient w.r.t. the generalized coordinate used for simulation i.e., $\partial \bm{C}(\bm{q})/\partial \bm{q}$ is constant or varying based on the current $\bm{q}$.

An ABD-based multibody system on the other hand, offers more flexibility. Because the CP coordinate $\bm{y}$ and the affine coordinate $\bm{q}$ are linearly related via Eq.~\eqref{eq:cp_coordinate}. The ball joint is a linear constraint. Similarly, a 6-DOF hinge constraint and a 12-DOF universal constraint are also linear since they are defined on the (global) CP coordinate $\bm{y}$. This has been considered a major advantage of ABD. When solving the multibody system in a primal way, the constant constraint Jacobian eases the implementation. For instance, when approximating joint constraints with quadratic penalty energies~\cite{UnifiedNewtonBarrier}, the penalty Hessian is constant. If we keep the rank of the hinge joint and universal joint minimal, the equality constraints are enforced at the local CP coordinate $\tilde{\bm{y}}$, which relates the global CP coordinate $\bm{y}$ via a varying rotation ($\bm{R}_H$, $\bm{R}_U$). That said, the 5-DOF hinge joint and the 4-DOF universal joint are nonlinear. 

In our implementation, we opt for the nonlinear version of the hinge and universal joints as of Eqs.~\eqref{eq:hinge} and \eqref{eq:universal} for more compact representations. This choice is closely related to how we would like to solve the corresponding multibody system. As to be detailed in Sec.~\ref{sec:dual_solve}, we design a primal-dual formulation that solves the system in the dual space, exploiting the re-factorized primal matrix of an affine body. A compact joint representation yields a smaller-sized dual matrix.

Normally, relative motions between two joint-linked bodies cannot be arbitrarily large. We combine the strain-limiting~\cite{provot1995deformation} and explicit penalty to tackle unilateral limits. Specifically, when we observe an out-of-range joint DOF $\theta$, we clamp it to its nearest valid setup $\theta \leftarrow \hat{\theta}$, and keep $\theta = \hat{\theta}$ for another Newton iteration while applying a dual-space penalty force as $k (\theta - \hat{\theta})$ at r.h.s. of Eq.~\eqref{eq:kkt}. In other words, we try to avoid any implicit mechanism for inequality joint constraints so that we can fully exploit the pre-factorized per-body Hessian.

\section{Dual-space KKT}\label{sec:dual_solve}
We now consider an ABD-based multibody system, consisting of $M$ bodies or links and $K$ joints. The corresponding KKT system at each Newton step is\footnote{The bottom part of the r.h.s. of Eq.~\eqref{eq:kkt} should be $-\widetilde{\bm{C}}(\bm{q}_n)$ i.e., the negative constraint residual from the previous iteration. Since our method often converges just with one iteration, this variable is always very small and thus ignored for simplicity. Adding $-\widetilde{\bm{C}}(\bm{q}_n)$ does not induce extra computational overhead.}:
\begin{equation}\label{eq:kkt}
   \begin{bmatrix}
   \widetilde{\bm{H}} & \nabla^\top \widetilde{\bm{C}} \\
   \nabla\widetilde{\bm{C}} & \bm{0}
   \end{bmatrix}
   \begin{bmatrix}
   \delta \widetilde{\bm{q}} \\
   \delta \widetilde{\bm{\lambda}}
   \end{bmatrix}
   =
   \begin{bmatrix}
   \widetilde{\bm{f}}_A\\
   \bm{0}
   \end{bmatrix},
\end{equation}
where $\widetilde{\bm{H}} = \operatorname{diag}_M(\bm{H}^j_A)$ is a block-diagonal matrix, whose $12 \times 12$ diagonal blocks are the reduced Hessian $\bm{H}_A^j$ i.e., Eq.~\eqref{eq:newton_abd}. The subscript $(\cdot)^j$is used to denote the body or joint index, and $\widetilde{(\cdot)}$ suggests the variable is global i.e., it stacks the local variables from all the bodies or joints. For instance, $\delta \widetilde{\bm{q}} \in \mathbb{R}^{12M}$ is the global primal unknown, and $\widetilde{\bm{f}}_A \in \mathbb{R}^{12M}$ is the global primal r.h.s. They concatenate the regular affine coordinate $\bm{q}^j$ and the affine force $\bm{f}_A^j$ of each body.  $\nabla \widetilde{\bm{C}}$ is the generalized constraint gradient such that $\nabla \widetilde{\bm{C}} = \partial \widetilde{\bm{C}} / \partial \widetilde{\bm{q}}$. $\delta \widetilde{\bm{\lambda}}$ is the dual DOFs or the Lagrange multipliers. Let $C_k$ be the rank of each joint. The dimensionality of $\delta \widetilde{\bm{\lambda}}$ is $\sum_{ k = 1}^K C_k$. 

Instead of using a minimal-coordinate formulation to solve Eq.~\eqref{eq:kkt}, we eliminate the primal variable and solve the system in the dual space. The reason is twofold. First, $\sum_{ k = 1}^K C_k$ is often a much smaller number than $12M$, making the dual version of Eq.~\eqref{eq:kkt} a much smaller problem. Second, the dual-space solve better exploits the co-rotated single-body ABD solver of Eq.~\eqref{eq:abd_final}. Expanding the first line of Eq.~\eqref{eq:kkt} yields:
\begin{equation}\label{eq:primal}
 \widetilde{\bm{H}} \delta \widetilde{\bm{q}} + \nabla^\top \widetilde{\bm{C}} \delta \widetilde{\bm{\lambda}} = \widetilde{\bm{f}}_A \Rightarrow \delta \widetilde{\bm{q}} = \widetilde{\bm{H}}^{-1} \left(\widetilde{\bm{f}}_A - \nabla^\top \widetilde{\bm{C}} \delta \widetilde{\bm{\lambda}} \right).
\end{equation}
Substituting Eq.~\eqref{eq:primal} back to the second line of Eq.~\eqref{eq:kkt} leads to:
\begin{equation}\label{eq:dual}
    \left( \nabla \widetilde{\bm{C}} \widetilde{\bm{H}}^{-1} \nabla^\top \widetilde{\bm{C}} \right) \delta \widetilde{\bm{\lambda}} = \nabla \widetilde{\bm{C}} \widetilde{\bm{H}}^{-1} \widetilde{\bm{f}}_A.
\end{equation}

\begin{wrapfigure}{r}{0.47\linewidth}
\includegraphics[width=\linewidth]{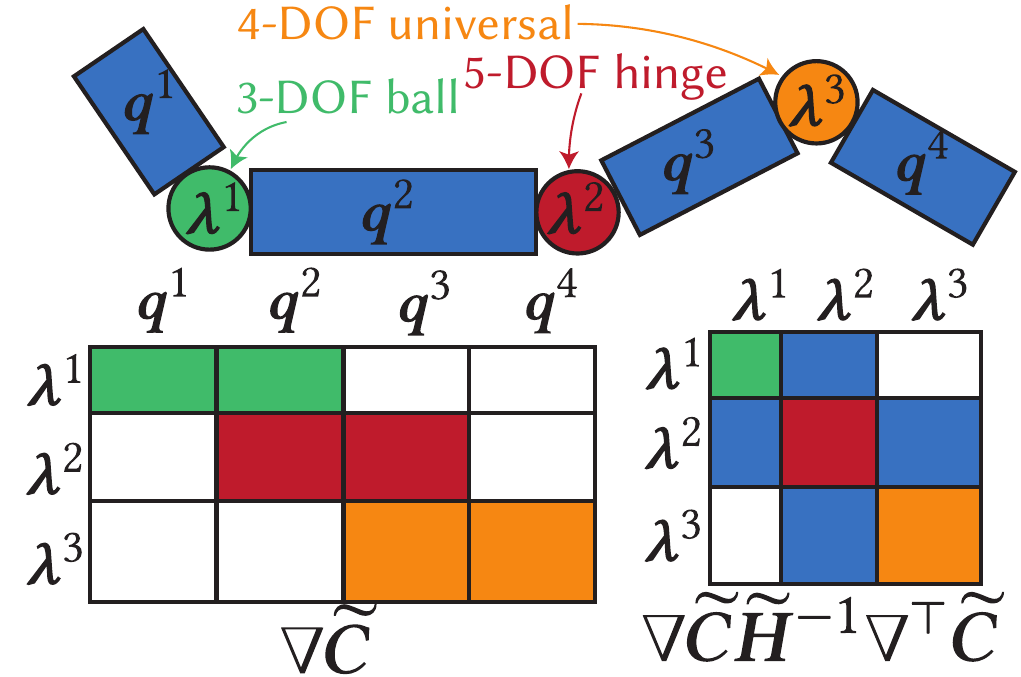}
\caption{\textbf{Block-sparse dual matrix.} The constraint gradient matrix $\nabla \widetilde{\bm{C}}$ is block-sparse, and the dual matrix is also block-sparse. An off-diagonal block is non-zero iff two joints are on the same affine body. }  
\label{fig:dual_mat}
\end{wrapfigure}
On the r.h.s. of Eq.~\eqref{eq:dual}, $\widetilde{\bm{H}}^{-1} \widetilde{\bm{f}}_A$ represents the constraint-free single ABD solve for all the bodies. It can be efficiently handled with Eq.~\eqref{eq:abd_final} i.e., Alg.~\ref{alg:abd}. The result is then projected to the tangent space of the constraint. 
The dual matrix $\nabla \widetilde{\bm{C}} \widetilde{\bm{H}}^{-1} \nabla^\top \widetilde{\bm{C}}$ is block-sparse. 
$\nabla \widetilde{\bm{C}}$ is also a sparse matrix as each joint only influences its incident affine bodies, and $\bm{\bm{H}}^{j}$ is pre-factorized. Therefore, $\widetilde{\bm{H}}^{-1} \nabla^\top \widetilde{\bm{C}}$ can be efficiently evaluated at each body. If all the joints are linear, meaning $\nabla \widetilde{\bm{C}}$ is constant, computing $\widetilde{\bm{H}}^{-1} \nabla^\top \widetilde{\bm{C}}$ using Eq.~\eqref{eq:abd_final} still needs to rotate $\nabla \bm{C}^k$ into the local frame attached at each affine body based on $\bm{q}^j$. Therefore, we care less about the linearity of a joint constraint but prefer having its rank minimized.  

\subsection{Constraint gradient}
Except for the ball constraint whose gradient is constant, the hinge, universal, and prismatic constraints can be uniformly written as:
\begin{equation}\label{eq:constriant}
    \bm{C}^k\left( \bm{R}_{Joint}(\bm{q}^\alpha, \bm{q}^\beta), \bm{q}^\alpha, \bm{q}^\beta \right) = 0,
\end{equation}
where $\bm{R}_{Joint}$ refers to the necessary rotation i.e. $\bm{R}_H$, $\bm{R}_U$, or $\bm{R}_P$ in Eqs.~\eqref{eq:hinge}, \eqref{eq:universal} and \eqref{eq:prismatic}, that converts $\bm{y}$ to $\tilde{\bm{y}}$. Here, we only consider two affine bodies $\alpha$ and $\beta$ connected by the $k$-th joint. The corresponding constraint gradient is:
\begin{equation}\label{eq:constraint_gradient}
    \nabla \bm{C}^k = \frac{\partial \bm{C}^k} {\partial \bm{R}_{Joint}} : \frac{\partial \bm{R}_{Joint}}{\partial \bm{q}^\alpha} + \frac{\partial \bm{C}^k} {\partial \bm{R}_{Joint}} : \frac{\partial \bm{R}_{Joint}}{\partial \bm{q}^\beta} + \frac{\partial \bm{C}^k}{\partial \bm{q}^\alpha} + \frac{\partial \bm{C}^k}{\partial \bm{q}^\beta}.
\end{equation}

Without loss of generality, we focus on the 5-DOF hinge constraint defined as Eq.~\eqref{eq:hinge}. In Eq.~\eqref{eq:constraint_gradient}, $\frac{\partial \bm{C}^k}{\partial \bm{q}^\alpha}$ and $ \frac{\partial \bm{C}^k}{\partial \bm{q}^\beta}$ are easy, which treat $\bm{R}_{Joint}$ as constant: 
\begin{equation}\label{eq:pcpq}
    \frac{\partial \bm{C}^k}{\partial \bm{q}^\alpha} = \bm{S}_H^\alpha \operatorname{diag}_4(\bm{R}_H) \bm{T}^\alpha, \quad \frac{\partial \bm{C}^k}{\partial \bm{q}^\beta} = -\bm{S}_H^\beta \operatorname{diag}_4(\bm{R}_H) \bm{T}^\beta.
\end{equation}
Similarly, the constraint derivatives w.r.t. the rotation when treating $\bm{q}^\alpha$ and $\bm{q}^\beta$ as constants are:
\begin{equation}\label{eq:pcpr}
    \frac{\partial \bm{C}^k}{\partial \operatorname{vec}(\bm{R}_H)} = 
\bm{S}^\alpha_H \Big (\operatorname{vec}^{-1}(\bm{T}^\alpha\bm{q}^\alpha)^\top \otimes \bm{I}_3 \Big ) - \bm{S}^\beta_H \Big (\operatorname{vec}^{-1}(\bm{T}^\beta\bm{q}^\beta)^\top \otimes \bm{I}_3 \Big ),
\end{equation}
where $\operatorname{vec}^{-1}$ reshapes $\bm{T}\bm{q}$ to a $3 \times 4$ matrix.

We also notice that $\bm{R}_H = \Delta \bm{R}^\alpha_H \bm{R}^{\alpha^\top}$ or $\bm{R}_H =  \Delta \bm{R}^\beta_H \bm{R}^{\beta^\top}$. That said, $\bm{R}_H$ can always be considered as a body-wise rotation converting the regular CP coordinate $\bm{y}$ to the local affine frame followed by $\Delta \bm{R}^\alpha_H$ (or $\Delta \bm{R}^\alpha_H$) to align a designated CT edge with the hinge axis. Both $\Delta \bm{R}^\alpha_H$ and $\Delta \bm{R}^\alpha_H$ are constant since the CTs are set. The rotation gradient (in the vectorized form) is then:
\begin{equation}\label{eq:r_gradient}
    \frac{\partial \operatorname{vec}(\Delta \bm{R}_H \bm{R}^\top)}{\partial \bm{q}} =(\bm{I}_3 \otimes \Delta \bm{R}_H)\; \frac{\partial\operatorname{vec}(\bm{R}^\top)}{\partial \bm{q}}. 
\end{equation}
Here, we further ignore the superscripts of $\alpha$ or $\beta$ for better clarity.  

An implicit (and accurate) constraint gradient evaluates the gradient of the rotation $\bm{R}$, which is cumbersome and often numerically unstable for an RBD-based multibody system. Fortunately, in ABD-based models, we always have $\bm{R} \approx \bm{A}$ and $\bm{q} = [\operatorname{vec}^\top(\bm{A}), \bm{t}^\top]^\top$ becomes a linear function of $\bm{A}$. Therefore, Eq.~\eqref{eq:r_gradient} can be efficiently estimated as:
\begin{equation}
     \frac{\partial \operatorname{vec}(\Delta \bm{R}_H \bm{R}^\top)}{\partial \bm{q}} \approx (\bm{I}_3 \otimes \Delta \bm{R}_H) \frac{\partial\operatorname{vec}(\bm{A}^\top)}{\partial \bm{q}} = (\bm{I}_3 \otimes \Delta \bm{R}_H) \tilde{\bm{I}}_{9 \times 12}.
\end{equation}
$\tilde{\bm{I}}_{9 \times 12}$ is a constant permutation matrix, which re-indexes entries in $\operatorname{vec}(\bm{A}^\top)$ to $\bm{q}$ with zero-padding at translation DOFs.

The approximate error of $\bm{R} \approx \bm{A}$ can also be easily eliminated by knowing the fact that  $\frac{\partial \operatorname{vec}(\Delta \bm{R}_H \bm{R}^\top)}{\partial \bm{q}}$ is the stacking of $12$ (i.e., the dimensionality of $\bm{q}$) vectorized skew-symmetric matrices.
Let $\bm{s}_\ell \in\mathbb{R}^9$ be the $\ell$-th row of $\frac{\partial \operatorname{vec}(\Delta \bm{R}_H \bm{R}^\top)}{\partial \bm{q}}$, we ``skew-symmetrize'' $\bm{s}_\ell$ by setting each of its elements as:
\begin{equation}\label{eq:skew}
\begin{aligned}
& s_{\ell, 1} \leftarrow 0,\quad s_{\ell, 5} \leftarrow 0,\quad s_{\ell, 9} \leftarrow 0, & (s_{\ell, 2}, s_{\ell, 4}) \leftarrow \pm \tfrac{1}{2}(|s_{\ell, 2}|+|s_{\ell, 4}|), \\
& (s_{\ell, 3}, s_{\ell, 7}) \leftarrow \pm \tfrac{1}{2}(|s_{\ell, 3}|+|s_{\ell, 7}|), &  (s_{\ell, 6}, s_{\ell, 8}) \leftarrow \pm \tfrac{1}{2}(|s_{\ell, 6}|+|s_{\ell, 8}|).
\end{aligned}
\end{equation}
The gradients of universal and prismatic joints can be obtained in a similar manner. 

From the above derivation, it should be noted that in an ABD-based multibody system, $\widetilde{\bm{C}}$ can be well approximated as a quadratic function of $\widetilde{\bm{q}}$ since $\bm{R}^j$ is relaxed to $\bm{A}^j$. As a result, $\nabla \widetilde{\bm{C}}$ becomes a linear function of $\widetilde{\bm{q}}$ i.e., see Eq.~\eqref{eq:pcpr}, which lumps multiple skew-symmetric matrices of $\frac{\partial \bm{R}_{Joint}}{\partial \bm{q}^j}$ into the constraint's tangent space.

In our implementation, we only keep three entries for each rotation gradient, and the corresponding computation is lightweight. Because $\widetilde{\bm{H}}$ is pre-factorized, building the dual matrix is also fast. At this point, we can use any linear solver like Cholesky decomposition to solve the dual matrix and convert the solution back to the primal variable via Eq.~\eqref{eq:primal}. Since the dual matrix is typically one-third or one-fourth the size of the primal system, we observe a significant speedup. Meanwhile, our framework enforces all the joints' constraints via KKT. Those constraints are well satisfied in most cases just using one Newton iteration.  

Articulated systems are often man-made and well-structured, which endows the dual matrices with specialized patterns. Next, we show how to leverage some common structures to optimize the computation. Our goal is to achieve efficient multibody simulation, even on a single CPU thread.

\subsection{Case I: Joint chain}\label{subsec:joint_chain}
Many multibody systems consist of a single kinematic chain (e.g., see Fig.~\ref{fig:dual_mat}), such as robotic arms. On such articulated structure, we have $M - K = 1$, and each joint couples exactly two bodies. Assume the $k$-th joint is $C_k$-DOF, and it couples bodies $j$ and $j+1$. On the kinematic chain, we can re-index joints and bodies such that $k = j$, and the corresponding constraint block becomes:
\begin{equation}
\nabla\bm C_j^j\,\delta \bm q^{j}+\nabla\bm C_{j+1}^{j}\, \delta \bm q^{j+1}=\bm 0,
\end{equation}
which leads to two dense blocks in the dual matrix:
\begin{align}\label{eq:chain-Schur-blocks}
\bm{D}^j &= \nabla\bm C_j^{j}(\bm H_A^j)^{-1}\nabla^\top\bm C_j^{j}
+
\nabla\bm C_{j+1}^{j}(\bm H_A^{j+1})^{-1}\nabla^\top\bm C_{j+1}^{j}
\ \in\mathbb R^{C_j\times C_j}, \nonumber\\
\bm{B}^j &= \nabla\bm C_{j+1}^{j}(\bm H_A^{j+1})^{-1}\nabla^\top\bm C_{j+1}^{j+1}
\ \in\mathbb R^{C_j\times C_{j+1}},
\end{align}
and the r.h.s. dense block is:
\begin{equation}
\bm b^j = 
\nabla\bm C_j^{j}(\bm H_A^j)^{-1}\bm f_A^{j}
+ \nabla\bm C_{j+1}^{j} (\bm H_A^{j+1})^{-1}\bm f_A^{j+1}\ \in\mathbb R^{C_j}.
\label{eq:chain-b}
\end{equation}
In other words, the dual matrix $\nabla \widetilde{\bm{C}} \widetilde{\bm{H}}^{-1} \nabla^\top \widetilde{\bm{C}}$ becomes \emph{block–tridiagonal} such that: 
\begin{equation}\label{eq:chain-btd}
\begin{bmatrix}
\bm D^1 & \bm B^1 & \\
\bm B^{1^\top} & \bm D^2 & \ddots \\
 & \ddots & \ddots & \bm B^{K-1}\\
 & & \bm B^{{K-1}^\top} & \bm D^K
\end{bmatrix}
\begin{bmatrix}
\delta\bm\lambda^1\\[3pt] \delta\bm\lambda^2\\[3pt] \vdots\\[3pt] \delta\bm\lambda^K
\end{bmatrix}
=
\begin{bmatrix}
\bm b^1\\[3pt] \bm b^2\\[3pt] \vdots\\[3pt] \bm b^K
\end{bmatrix}.
\end{equation}
It is known that such a block-tridiagonal system can be solved with $O(K)$ block-wise eliminations using a block Thomas algorithm~\cite{press2007numerical}.
$\{\delta\bm\lambda_k\}_{k=1}^{K}$ are known. 
Specifically, for each body $j$, we have
\begin{equation}
\bm H_A^{j}\,\delta\bm q^{j}
+
\nabla^\top \bm C_{j}^{j-1} \,\delta\bm\lambda^{j-1}
+
\nabla^\top \bm C_{j}^{j}\,\delta\bm\lambda^{j}
=
\bm f_A^{j}.
\label{eq:recover-dq-perbody-stationarity}
\end{equation}
Therefore, each body's update boils down to the explicit form of:
\begin{equation}
\delta\bm q^{j}
=
\bm H_A^{{j}^{-1}}
\Big(
\bm f_A^{j}
-
\nabla^\top \bm C_{j}^{j-1}\,\delta\bm\lambda^{j-1}
-
\nabla^\top \bm C_{j}^{j}\,\delta\bm\lambda^{j}
\Big).
\label{eq:recover-dq-explicit}
\end{equation}

\begin{algorithm}
\DontPrintSemicolon
\KwIn{$\{\bm H_A^{j},\bm q^{j},\dot{\bm q}^{j},\bm M_A^{j},\bm W_{ext}^{j},\bm S^{j}\}_{j=1}^M,\ h$}
\KwOut{$\{\delta \bm q^{j}\}_{j=1}^M$}
\BlankLine

\For{$j=1$ \KwTo $M$}{
$\bm A^{j} \gets \bm A(\bm q^{j})$\;
$\bm G^{j} \gets \bm G(\bm A^{j})$\;
$\bm f_{A, ext}^{j} \gets \bm G^{j^\top}\bm W_{ext}^{j}$\;
$\bm f_A^{j} \gets \bm f_{A,ext}^{j} + \tfrac{1}{h}\bm M_A^{j}\dot{\bm q}^{j}$\;
$\bar{\bm H}_A^{j} \gets \bm H_A^{j}$,\quad $\bar{\bm f}_A^{j} \gets \bm f_A^{j}$\;
}

\For{link $j$ (leaf-to-root)}{
$p \gets \operatorname{parent}(j)$\;
$\bm S_{abd}^{j} \gets \bm E(\bm A^{j})\,\bm S^{j}$\;
$\bm U^{j} \gets \hat{\bm H}_A^{j}\,\bm S_{abd}^{j}$,\quad$\bm D^{j} \gets \bm S_{abd}^{j\top}\bm U^{j}$\;
solve $\bm\alpha^{j}$ via $\bm D^{j}\bm\alpha^{j}=\bm S_{abd}^{j^\top}\hat{\bm f}_A^{j}$\;
$\Delta\bm H_A^{j} \gets \bar{\bm H}_A^{j}-\bm U^{j}(\bm D^{j})^{-1}\bm U^{j\top}$,\quad$\Delta\bm f_A^{j} \gets \hat{\bm f}_A^{j}-\bm U^{j}\bm\alpha^{j}$\;
$\bm R_{\mathrm{rel}}^{j} \gets \bm A^{p\top}\bm A^{j}$\;
$\bm T^{j} \gets \mathrm{diag}_4(\bm R_{\mathrm{rel}}^{j^\top})$\;
$\hat{\bm H}_A^{p} \gets \hat{\bm H}_A^{p} + \bm T^{j^\top}\Delta \bm H_A^{j}\bm T^{j}$,\quad$\hat{\bm f}_A^{p} \gets \hat{\bm f}_A^{p} + \bm T^{j^\top}\Delta\bm f_A^{j}$\;
}

\For{link $j$ (root-to-leaf)}{
$p \gets \operatorname{parent}(j)$\;
$\bm r^{j} \gets -\nabla\bm C_{p}^{j}\,\delta\bm q^{p}$\;
solve $(\delta\bm q^{j},\delta\bm\lambda^{j})$ via Eq.~\eqref{eq:module3-kkt}
}
\caption{ABD Articulated Body Algorithm (ABD-ABA)}\label{alg:aba}
\end{algorithm}

\subsection{Case II: Joint tree (ABD-ABA)}\label{subsec:tree}
A generalization of a chain of joints is a tree of joints --- the whole articulated system forms a kinematic hierarchy such that each body has exactly one parent, and the system has exactly one root body. Such a structure allows us to solve the kinematic propagation level by level. This is known as the Articulated Body Algorithm (ABA) or Featherstone algorithm~\cite{featherstone2008rigid}. We show that this strategy can be conveniently (and more efficiently) transplanted to an ABD-based multibody system, and we name the resulting algorithm ABD-ABA. The pseudo code is listed in Alg.~\ref{alg:aba}. Similar to the classic ABA, the core idea is to treat each link as a pre-computed quadratic model in ABD increments and to propagate subtree contributions through a structured condensation, so that the dominant cost depends on the kinematic connectivity rather than the full state dimension.

For the link $j$, we denote its velocity as 
$\dot{\bm q}^j = [\dot{\bm q}_1^{j^\top},\dot{\bm q}_2^{j^\top},\dot{\bm q}_3^{j^\top},\dot{\bm t}^{j^\top}]^\top$ and build a linear map $\bm G(\bm A^j)\in\mathbb R^{6\times12}$ that converts ABD velocity to the spatial twist
$\bm V^j=[\boldsymbol\omega^{j^\top},\ \bm v^{j^\top}]^\top$:
\begin{equation}
\bm G(\bm A^j)=
\begin{bmatrix}
\tfrac12[\bm q_{1}^j]_\times & \tfrac12[\bm q_{2}^j]_\times & \tfrac12[\bm q_{3}^j]_\times & \bm 0_{3\times 3}\\[2pt]
\bm 0_{3\times 3} & \bm 0_{3\times 3} & \bm 0_{3\times 3} & \bm I_{3}
\end{bmatrix},\quad
\bm V^j = \bm{G}\dot{\bm q}^j.
\label{eq:module1-G-twist}
\end{equation}
Equivalently, it yields:
\begin{equation}
\bm v^j=\dot{\bm t}^j,\quad
\bm\omega^j=\frac12\Big(\bm q_{1}^j\times\dot{\bm q}_{1}^j+
\bm q_{2}^j\times\dot{\bm q}_{2}^j+\bm q_{3}^j\times\dot{\bm q}_{3}^j\Big).
\label{eq:module1-omega-v}
\end{equation}
Given an external spatial wrench $\bm W^j_{ext}=[\bm\tau^{j^\top}_{ext},\ \bm f_{ext}^{j^\top}]^\top$, the corresponding affine force in ABD coordinates can be derived following the principle of virtual work i.e., $\bm f_{A, ext}^{j^\top}\dot{\bm q}^j=\bm W_{ext}^{j\top}\bm V^j$ such that:
\begin{equation}\label{eq:module1-wrench-map}
f_{A}^j = f_{A, ext}^j + \frac{1}{h}\bm M_{A}^j\dot{\bm q}^j, \quad
\bm f_{A, ext}^j = \bm G(\bm A^j)^\top\,\bm W_{ext}^j.
\end{equation}
To see why we do not need an explicit gyroscopic term, we start with the equation of motion of RBD:
\begin{equation}
\bm W^j_{dyn} = \bm I^j \dot{\bm V}^j + \bm V^j\times^\ast(\bm I^j\bm V^j),
\label{eq:spatial-dyn}
\end{equation}
where $\bm I^j \in \mathbb{R}^{6 \times 6}$ is the spatial inertia of the rigid body. $\times^\ast$ is the force cross operator. We apply time differentiation at both side of $\bm V^j = \bm G(\bm A^j)\dot{\bm q}^j$, which gives
$\dot{\bm V}^j = \bm G\,\ddot{\bm q}^j + \dot{\bm G}\,\dot{\bm q}^j$. We can then project Eq.~\eqref{eq:spatial-dyn} to the affine coordinate such that:
\begin{equation}\label{eq:projected-spatial}
\bm f_{A, dyn}^j
= \underbrace{\bm G^\top \bm I^j \bm G}_{\bm M_A^j}\,\ddot{\bm q}^j
+ \Big( \bm G^\top \bm I^j \dot {\bm G}\,\dot{\bm q}^j +
\bm G^\top\!\big(\bm V^j\times^\ast(\bm I^j\bm V^j)\big) \Big).
\end{equation}
On the other hand, the kinetic energy in ABD coordinates is $T^j=\tfrac12\,\dot{\bm q}^{j^\top}\bm M_A^j \dot{\bm q}^j$. The Euler-Lagrange equation can be written as:
\begin{equation}
\bm f_{A, dyn}^j
=
\bm M_A^j\ddot{\bm q}^j
+
\Big(\dot{\bm M}_A^j\,\dot{\bm q}^j-\frac{\partial T^j}{\partial \bm q^j}\Big).
\label{eq:EL}
\end{equation}
From Eqs.~\eqref{eq:projected-spatial} and \eqref{eq:EL}, we can find the following correspondence:
\begin{equation}
\bm{G}^\top \bm I^j \dot{\bm{G}}\,\dot{\bm q}^j
+
\bm{G}^\top\!\big(\bm V^j\times^\ast(\bm I^j\bm V^j)\big)
=
\dot{\bm M}_A^j\,\dot{\bm q}^j-\frac{\partial T^j}{\partial \bm q^j}.
\label{eq:cancellation-identity}
\end{equation}
In ABD, $\bm M_A^j$ is constant and $T^j$ does not depend on $\bm q^j$,
so $\dot{\bm M}_A^j=\bm 0$ and $\frac{\partial T^j}{\partial \bm q^j} =\bm 0$. Therefore, we have:
\begin{equation}\label{eq:cancellation-zero}
\bm{G}^\top \bm I^j \dot{\bm{G}} \,\dot{\bm q}^j
+ \bm{G}^\top\!\big(\bm V^j\times^\ast(\bm I^j\bm V^j)\big) =\bm 0.
\end{equation}
That said, the projected spatial bias wrench exactly cancels the contribution from $\dot{\bm{G}}$. Consequently, the gyroscopic term is not needed in ABD equations of motion.

During the upward condensation, each local matrix and force is accumulated sequentially 
such that i.e., $\hat{\bm H}_{A}^j \gets \bm H_{A}^j$, and $\hat{\bm f}_A^j \gets \bm f^j_A$. We first perform the leaf-to-root recursion within ABD coordinates. In standard Featherstone ABA, each joint $k$ has a subspace matrix $\bm S^k\in\mathbb R^{6\times m_k}$, which maps joint velocity to the relative spatial twist by $\bm V_{rel}=\bm S^k\,\dot{\bm \theta^k}$. Here, $m_k = 6-C_k$ gives the joint's DOF count.  To express the same admissible motion in with ABD coordinates, we use the rigid-motion embedding of $\bm E(\bm A^j)\in\mathbb R^{12\times 6}$ such that:
\begin{equation}
\bm E(\bm A^j)= \begin{bmatrix}
-[\bm q_{1}^j]_\times & \bm 0_{3\times3}\\
-[\bm q_{2}^j]_\times & \bm 0_{3\times3}\\
-[\bm q_{3}^j]_\times & \bm 0_{3\times3}\\
\bm 0_{3\times3} & \bm I_3
\end{bmatrix},
\qquad
\dot{\bm q}^j = \bm E(\bm A^j)\,\bm V^j.
\label{eq:module2-E}
\end{equation}
This embedding is consistent with the twist map: for rigid motions,
$\bm V = \bm G(\bm A)\dot{\bm q}$ and $\bm G(\bm A)\bm E(\bm A)=\bm I_6$.

As all the bodies are tree-structured, we also have $M - K = 1$ and can re-index joints and bodies such that $k = j$. The ABD joint subspace simply becomes:
\begin{equation}
\bm S_{abd}^j
=
\bm E(\bm A^j)\,\bm S^j
\in\mathbb R^{12\times m_j}.
\label{eq:module2-Sabd}
\end{equation}
Given $\bm S^j$, we eliminate the joint DOFs of link $j$ via a condensation analogous to articulated inertia recursion. This is achieved by assembling $\bm U^j = \hat{\bm H}_{A}^j\bm S^j_{abd}$, $\bm D^j = \bm S^{j^\top}_{abd} \bm U^j \in \mathbb R^{m_j\times m_j}$, and solving:
\begin{equation}
\bm D^j\,\bm\alpha^j = \bm S^{j^\top}_{abd} \hat{\bm f}_A^j.
\label{eq:module2-alpha}
\end{equation}
The condensed contributions of subtree $j$ (with joint coordinates eliminated) are:
\begin{equation}
\Delta\bm H_{A}^j = \hat{\bm H}_{A}^j - \bm U^j\,(\bm D^{j})^{-1}\,\bm U^{j^\top},
\quad
\Delta\bm f^j_A = \hat{\bm f}^j_A - \bm U^j\,\bm\alpha^j.
\label{eq:module2-condense}
\end{equation}
Here, $\Delta\bm H_{A}^j$ and $\Delta\bm f^j_A$ encode how the subtree rooted at $j$ reacts to a parent-frame perturbation once the joint DOFs at $(p,j)$ are internally resolved, where $p$ is the index of the parent of body $j$. 


Because our ABD increment stacks four $3$-vectors (three columns of $\bm A$ plus $\bm t$), and the relative rotation is $\bm R_{rel}=\bm A^{p\top} \bm A^j$, we use the block-diagonal transform $\bm X^j = \operatorname{diag}_4(\bm R_{rel}^\top)$
to rotate the condensed contribution into the parent coordinates, and accumulate:
\begin{equation}
\hat{\bm H}_{A}^p \mathrel{+}= \bm X^{j^\top} \Delta\bm H_{A}^{j}\bm X^j,
\quad
\hat{\bm f}_A^p \mathrel{+}= \bm X^{j^\top} \Delta\bm f^j_A.
\label{eq:module2-accumulate}
\end{equation}

After the upward pass, each link $j$ stores an articulated (condensed) pair of
$(\hat{\bm H}_{A}^j,\hat{\bm f}_A^j)$ that summarizes the effect of the entire subtree rooted at $j$. We then perform a root-to-leaf traversal to compute the per-link ABD increment $\delta\bm q^j$ while enforcing the joint constraints at each parent-child attachment.

Let $\bm C^j(\bm q^p,\bm q^j)=\bm 0$ be the joint constraint between the parent $p$ and child $j$. Linearizing at the current configuration yields:
\begin{equation}
\nabla\bm C^{j}_{p}\,\delta\bm q^p
+
\nabla\bm C^{j}_{j}\,\delta\bm q^j
=\bm 0.
\label{eq:module3-linearize-full}
\end{equation}
Given that $\delta\bm q^p$ is already known in the downward pass, we rewrite it as a linear equality in the child increment:
\begin{equation}
\nabla\bm C^{j}_{j}\,\delta\bm q^j = \bm r^j,
\qquad
\bm r^j = -\nabla\bm C^{j}_{p}\,\delta\bm q^p,
\label{eq:module3-di}
\end{equation}
where $\nabla\bm C^{j}_{p}$ and $\nabla\bm C^{j}_{j}$ are in the same coordinate frame as $\delta\bm q^j$.

The increment $\delta\bm q^j$ is obtained by solving the local KKT of:
\begin{equation}\label{eq:module3-kkt}
\begin{bmatrix}
\hat{\bm H}_{A}^j & \nabla\bm C_{j}^{j^\top}\\
\nabla\bm C_{j}^{j} & \bm 0
\end{bmatrix}
\begin{bmatrix}
\delta\bm q^j\\
\delta\bm\lambda^j
\end{bmatrix}
=
\begin{bmatrix}
\hat{\bm f}_A^j\\
\bm r^j
\end{bmatrix}.
\end{equation}
Similar to its global counterpart i.e., Eq.~\eqref{eq:kkt}, we eliminate the primal unknown $\delta\bm q^j$, and solve for the multiplier first. As a result, the global nonlinear solve is decomposed into a series of lightweight joint-size local solves.

\subsection{Case III: Joint loop}
A joint loop is made by a chain of bodies whose first and last bodies are also constrained by a joint. In other words, it rolls back to a joint chain if we remove any affine body from the system. 

Let $M$ be the body we temporarily remove, and it is constrained by joints $K$ and $1$. We partition the global KKT of Eq.~\eqref{eq:kkt} into two blocks of DOFs. The first block contains all the primal DOFs from body $j = 1$ to $j = M - 1$, as well as all the dual DOFs from joints $k = 2$ to $k = K-1$. The body $M$ and pertaining joints $K$ and $1$ are in the other block such that:
\begin{equation}\label{eq:loop}
   \begin{bmatrix}
   \mathcal{A} & \mathcal{C}^\top \\
   \mathcal{C} & \mathcal{D}
   \end{bmatrix}
   \begin{bmatrix}
   \bm{w}_{\mathcal{A}} \\
   \bm{w}_{\mathcal{D}}
   \end{bmatrix}
   =
   \begin{bmatrix}
  \bm{b}_{\mathcal{A}}\\
   \bm{b}_{\mathcal{D}}
   \end{bmatrix},
\end{equation}
where $\mathcal{A}$ represents the system of the joint chain, which can be converted to Eq.~\eqref{eq:chain-btd} and efficiently solved with block Thomas. Therefore, we solve Eq.~\eqref{eq:loop} by computing the Schur complement $\mathcal{S} = \mathcal{D} - \mathcal{C} \mathcal{A}^{-1} \mathcal{C}^\top$, which is a low-rank system, and substituting the $\bm{w}_\mathcal{D}$ back to the first line of Eq~\eqref{eq:loop} to retrieve unknowns in the first block.

This idea can be further generalized for loops on a joint tree. In this case, $\mathcal{A}$ represents an ABD-ABA system. As long as the total number of loops is limited, we can always use Schur complement to solve the sub-system corresponding to DOFs at the ``loop breakers''.

\subsection{Case IV: Joint graph}
The most generic situation is when the joint topology of the multibody system forms a graph. In this case, we propose a multi-directional block Gauss-Seidel procedure if a direct matrix factorization is prohibitive in a single-thread simulation environment. The high-level idea is to view the system as the combination of multiple joint chains, and we relax the strain in the dual space joint by joint, following those pre-defined chains.
\begin{table}
\centering
\footnotesize 
\caption{\textbf{Experiment statistics.} For each experiment shown in the paper, we report the number of links (\textbf{\# Link}), the total number of Constraints (\textbf{\# Cons.}), the time step size $h$, the constraint residual tolerance $\|\bm r\|$, the number of iterations per step (\textbf{\#~Iter.}), the Young's modulus of affine body $E$, and the per-step simulation time (\textbf{Sim.}) in milliseconds.}
\label{tab:stats}

\setlength{\tabcolsep}{1.0pt}
\renewcommand{\arraystretch}{1.15}

\begin{tabularx}{\columnwidth}{
    >{\raggedright\arraybackslash}X |
    >{\centering\arraybackslash}p{0.09\columnwidth} 
    >{\centering\arraybackslash}p{0.10\columnwidth} 
    >{\centering\arraybackslash}p{0.08\columnwidth} 
    >{\centering\arraybackslash}p{0.08\columnwidth} 
    >{\centering\arraybackslash}p{0.09\columnwidth} 
    >{\centering\arraybackslash}p{0.06\columnwidth} 
    >{\centering\arraybackslash}p{0.08\columnwidth}
}
\hline
\textbf{Scene} & \textbf{\# Link} & \textbf{\# Cons.} & $h$ & $\|\bm r\|$ & \textbf{\# Iter.} & $E$ &\textbf{Sim.} \\
\hline
Joint net (Fig.~\ref{fig:bignet_cylinder}: $100\times100$)  & 30K  & 120K  & $10~ms$ & 1E-7 & 1 & 1E9 & $84~ms$\\
A pulley system (Fig.~\ref{fig:chain})                      & 1.5K & 4.5K & $10~ms$ & 1E-7 & 1 & 1E8  & $2~ms$\\
A huge pulley system (Fig.~\ref{fig:huge_chain})            & 1M   & 3M  & $10~ms$ & 1E-5 & 1 & 1E8  & $904~ms$\\
Willow (Fig.~\ref{fig:tree}: top)                           & 21K  & 63K  & $10~ms$ & 1E-6 & 1 & 1E8  & $18~ms$\\
Pear (Fig.~\ref{fig:tree}: bottom)                          & 29K  & 87K  & $10~ms$ & 1E-6 & 1 & 1E8  & $23~ms$\\
Net cloak (Fig.~\ref{fig:net_cloth})                        & 12K  & 48K  & $10~ms$ & 1E-7 & 1 & 1E9 & $33~ms$\\
Armadillo (Fig.~\ref{fig:armadillo})                        & 2.7K  & 10.8K  & $10~ms$ & 1E-6 & 7 & 1E6 & $116~ms$\\
Ragdolls (Fig.~\ref{fig:ragdolls})                          & 1.5K & 6K  & $5~ms$  & 1E-6 & 1 & 1E6  & $17~ms$\\
Falling joints  (Fig.~\ref{fig:falling_joints})             & 1.4K   & 2.9K   & $10~ms$ & 1E-6 & 1 & 1E9  & $54~ms$\\
Protein (Fig.~\ref{fig:protein})                            & 14K  & 56K  & --   & 1E-7 & 1 & 1E6 & $14~ms$  \\
\hline
\end{tabularx}
\end{table}

\section{Experimental Results}
We implemented the M-ABD framework on a desktop equipped with an \texttt{AMD} \texttt{9950X3D} CPU, using Eigen~\cite{eigenweb} and MKL~\cite{wang2014intel} for linear algebra operations. We specifically target a single-thread CPU implementation, as multibody simulations in tasks like large-scale data synthesis for robotic training are typically parallelized across multiple independent instances. For complex systems with massive joint networks, the dual solver can be further accelerated via GPU parallelization. Nevertheless, this is not the scope of this paper. The statistics of the experimental setup and timing information is reported in Tab.~\ref{tab:stats}. Please refer to the supplementary video for additional animation results.

\begin{figure}
\centering
\includegraphics[width=\linewidth]{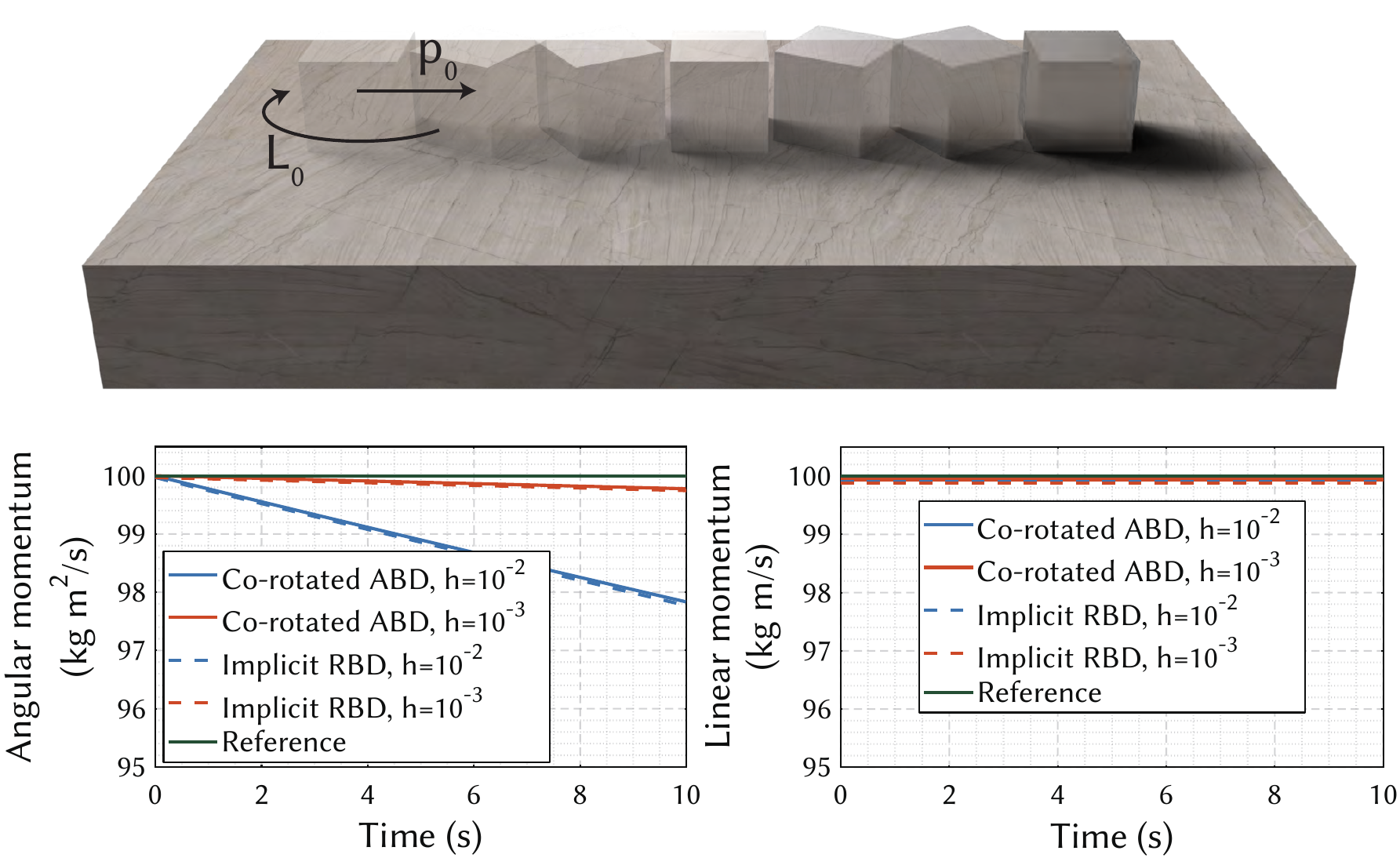}
\caption{\textbf{Spinning box.} 
We compare ABD (solid curves) against an implicit RBD baseline (dashed curves) for a cube with initial $\bm p_0=[100, 0, 0]^\top~{kg\,m/s}$ and $\bm L_0=[0, 100, 0]^\top~{kg\,m^2/s}$. Co-rotated ABD formulation closely matches the behavior of an implicit RBD.}
\label{fig:cube}
\end{figure}

\subsection{Single-body comparison}
We first quantitatively analyze the accuracy and reliability of the co-rotated ABD model for single stiff object simulation. We monitor the linear momentum, angular momentum, and total energy to validate that our method matches expected physical results under several standard benchmarks.

In Fig.~\ref{fig:cube}, we compare our co-rotated ABD with a standard implicit RBD baseline using a single cube moving on a frictionless surface. The size of the cube is $0.1~m$ ($\rho=10^3 \text{ kg/m}^3, E=10^9 \text{ Pa}, \nu=0.3$) with initial momenta $\bm p_0=[100, 0, 0]^\top$ and $\bm L_0=[0, 100, 0]^\top$. To initialize the motion, we compute the target spatial twist $\bm V_0 = [\bm\omega_0^\top, \bm v_0^\top]^\top$ and map it to ABD generalized velocities via Eq.~\eqref{eq:module1-G-twist}. Similarly, external spatial wrenches $\bm W_{ext} = [\bm{\tau}_{ext}^\top, \bm f_{ext}^\top]^\top$ are converted to ABD generalized forces $\bm f_{A, ext}$ using Eq.~\eqref{eq:module1-wrench-map}. Following the principle of virtual work, the equivalent spatial wrench can be obtained as:
\begin{equation}
\bm W_{ext}=\bm E(\bm A)^\top \bm f_{A,ext}.
\end{equation}
The variations of the momentum are plotted in the figure. We observe that linear momentum is preserved for all step sizes, while angular momentum shows dissipation that increases with $h$. Our ABD method matches the RBD baseline, and the conservation improves as $h$ decreases. This confirms that affine linearization does not violate the D'Alembert principle. The error mainly comes from the numerical damping of the time integrator. 

\begin{figure}
\centering
\includegraphics[width=\linewidth]{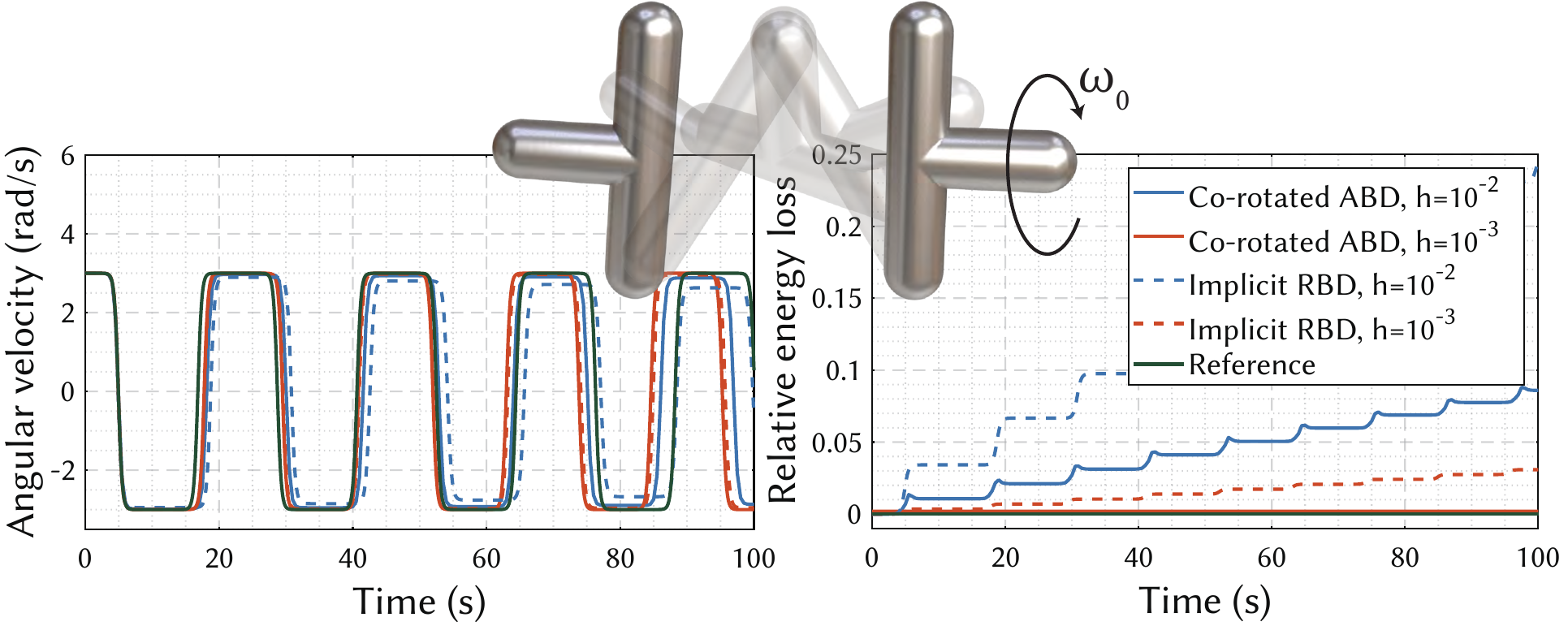}
\caption{\textbf{T-handle.}
Our co-rotated ABD integrator accurately reproduces the intermediate-axis instability per the intermediate-axis theorem. Left: We plot the body-frame angular velocity along the intermediate principal axis, which undergoes periodic sign flips. Right: We plot the energy loss over time, which increases during the rapid flipping events and remains nearly constant during the quasi-steady spinning phases between flips.}
\Description{}
\label{fig:T-handle}
\end{figure}
\begin{figure}
\centering
\includegraphics[width=\linewidth]{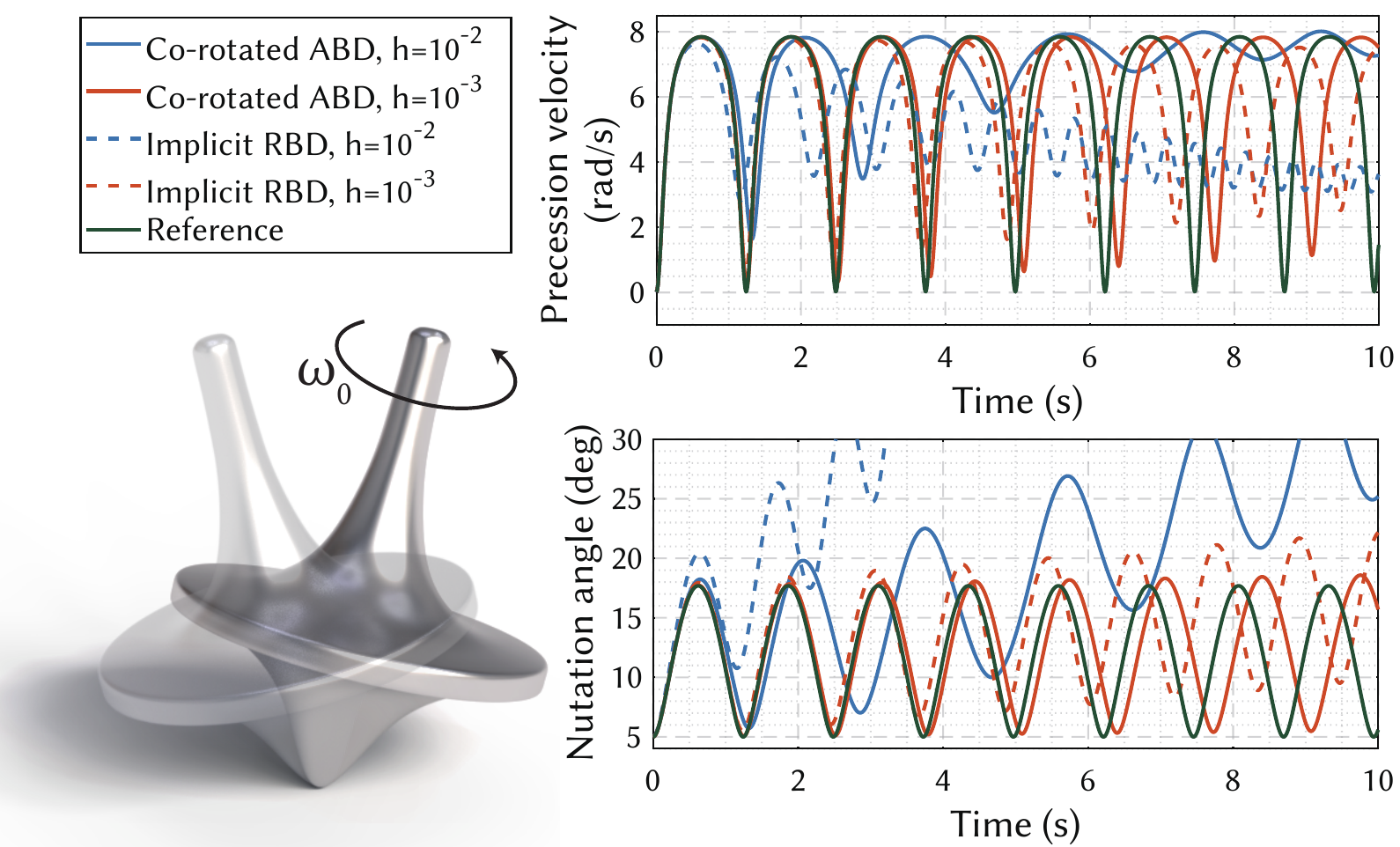}
\caption{\textbf{Heavy top.} 
We simulate a standard heavy top with a fixed pivot under gravity. The model is initialized with an angular speed of $10~{rad/s}$ and a $5^\circ$ tilt. We plot the precession velocity and the nutation angle over time using our method and implicit RBD under different time step sizes. Our method captures the expected coupling between spin and nutation, and the slight inertia asymmetry induces small oscillations in the nutation angle.}
\Description{}
\label{fig:spinning_top}
\end{figure}
In another benchmark shown in Fig.~\ref{fig:T-handle}, we use a T-handle model to evaluate how well the co-rotated ABD formulation captures the coupling among rotational DOFs. In a zero-gravity environment, we initialize the body with a nonzero angular velocity aligned with its intermediate principal axis. According to the intermediate-axis theorem~\cite{goldstein1950classical}, the T-handle should undergo periodic flipping. We set the initial angular speed to $\omega_0=3~{rad/s}$ and introduce a subtle asymmetry in the mass distribution. The motion is simulated using various time step sizes and compared against the reference, which is the result obtained with the implicit RBD solver using fourth-order Runge-Kutta (RK4) integration at $h=10^{-4}~s$. Similar to Fig.~\ref{fig:spinning_top}, the co-rotation ABD formula closely matches implicit RBD simulations. Yet, our method is over $30\%$ faster. 

In Fig.~\ref{fig:spinning_top}, we test a heavy top with a fixed pivot to see if our ABD method captures classic gyroscopic effects like precession and nutation. The motion is driven by gravity, with the top initialized with a $5^\circ$ tilt and an angular speed of $10~rad/s$. As in Fig~\ref{fig:T-handle}, the reference is from a high-accuracy RK4 integration ($h=10^{-4}~s$). ABD accurately reproduces the expected motion and converges to the reference as $h$ decreases. Notably, ABD shows less sensitivity to the time step size than the implicit RBD baseline, which suffers from more significant waveform distortion at coarser steps.

\begin{figure}
\centering
\includegraphics[width=\linewidth]{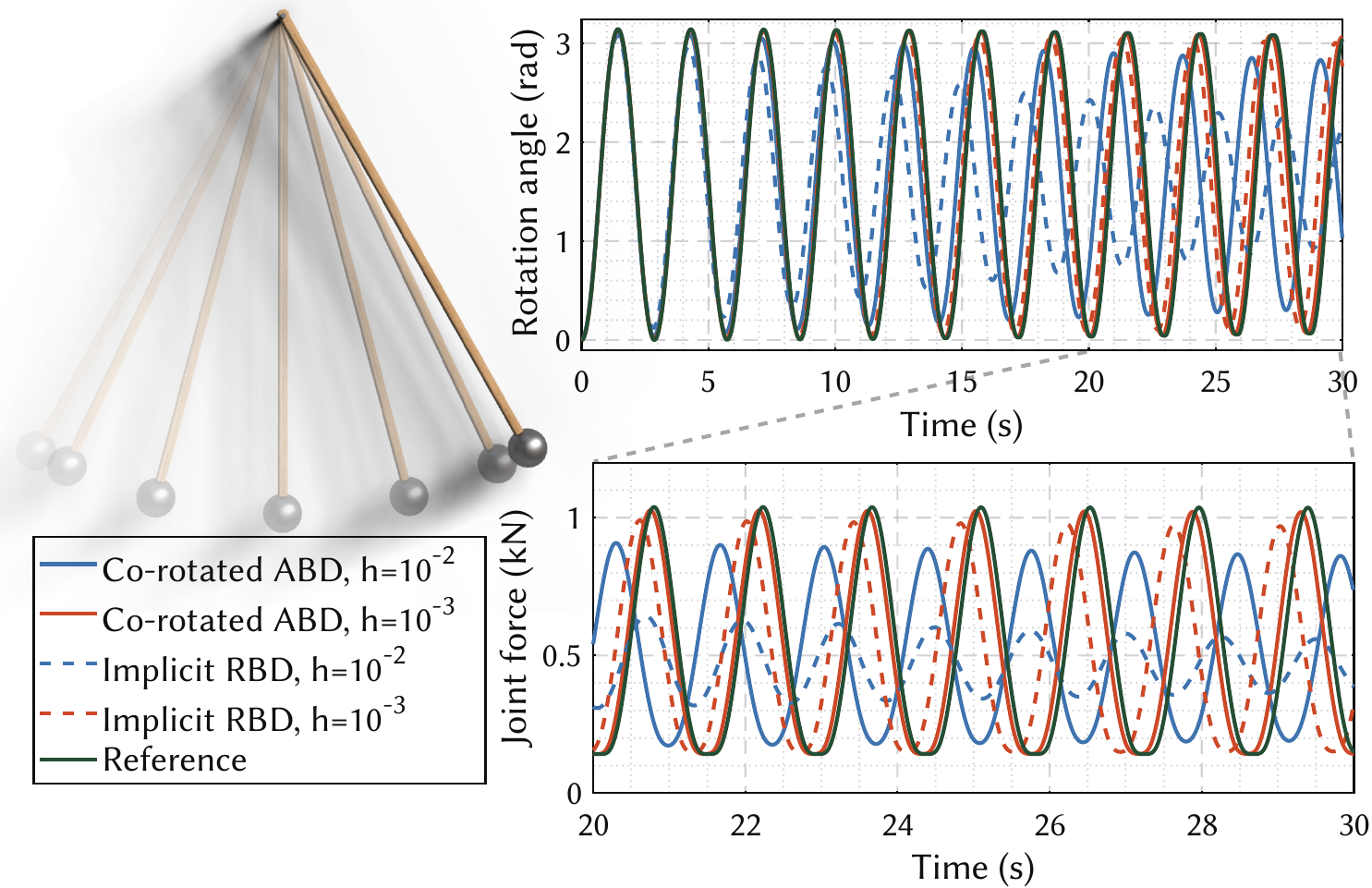}
\caption{\textbf{Physical pendulum.} 
We simulate a physical pendulum with a fixed pivot, released from a horizontal configuration with zero initial velocity under gravity. Top: We plot the pendulum angle over time, where the reference curve is from the elliptic-integral (Eq.~\ref{eq:elliptic}). Bottom: We plot the magnitude of the joint force over time, which varies periodically and peaks near the turning points of the swing. As the time step decreases, both the angle trajectory and the joint-force waveform produced by ABD better match the reference. Larger time steps lead to accumulated phase drift.}
\Description{}
\label{fig:pendulum}
\end{figure}

In Fig.~\ref{fig:pendulum}, we simulate a physical pendulum starting from a horizontal position. The pendulum swings freely under gravity, and we compare our ABD method and an implicit RBD baseline against the analytic solution based on elliptic integrals of:
\begin{equation}\label{eq:elliptic}
\theta(t)=\frac{\pi}{2}-2\,\arcsin\!\Big(\kappa\,\mathrm{sn}\!\big(K(\kappa)-\omega_{\mathrm{lin}} t,\kappa\big)\Big),
\end{equation}
where $K(\kappa)$ is the complete elliptic integral of the first kind and $\mathrm{sn}(\cdot,\kappa)$ is the Jacobi elliptic sine.
Our method tracks the analytic reference more closely than the implicit RBD baseline and provides more stable results for both the motion and joint forces.

\subsection{Multi-body comparison}
We now test our method on various multibody systems, focusing on constraint satisfaction and energy loss. Our primary competitors include MuJoCo~\cite{todorov2012mujoco}, Bullet~\cite{coumans2015bullet}, and PhysX~\cite{nvidia2025physx} which are popular off-the-shelf articulated body simulation solutions. Another important peer technique is quaternion-based multibody system or (VQ)~\cite{VersatileQuaternion}. VQ uses quaternion to parameterize the rotational motion of a rigid body, and it solves the full articulated dynamics with Newton's method. VQ also exactly enforces the joint constraints as our method. Nevertheless, this method is less scalable and becomes prohibitive as the total number of body increases.   

\begin{figure}
\centering
\includegraphics[width=0.9\linewidth]{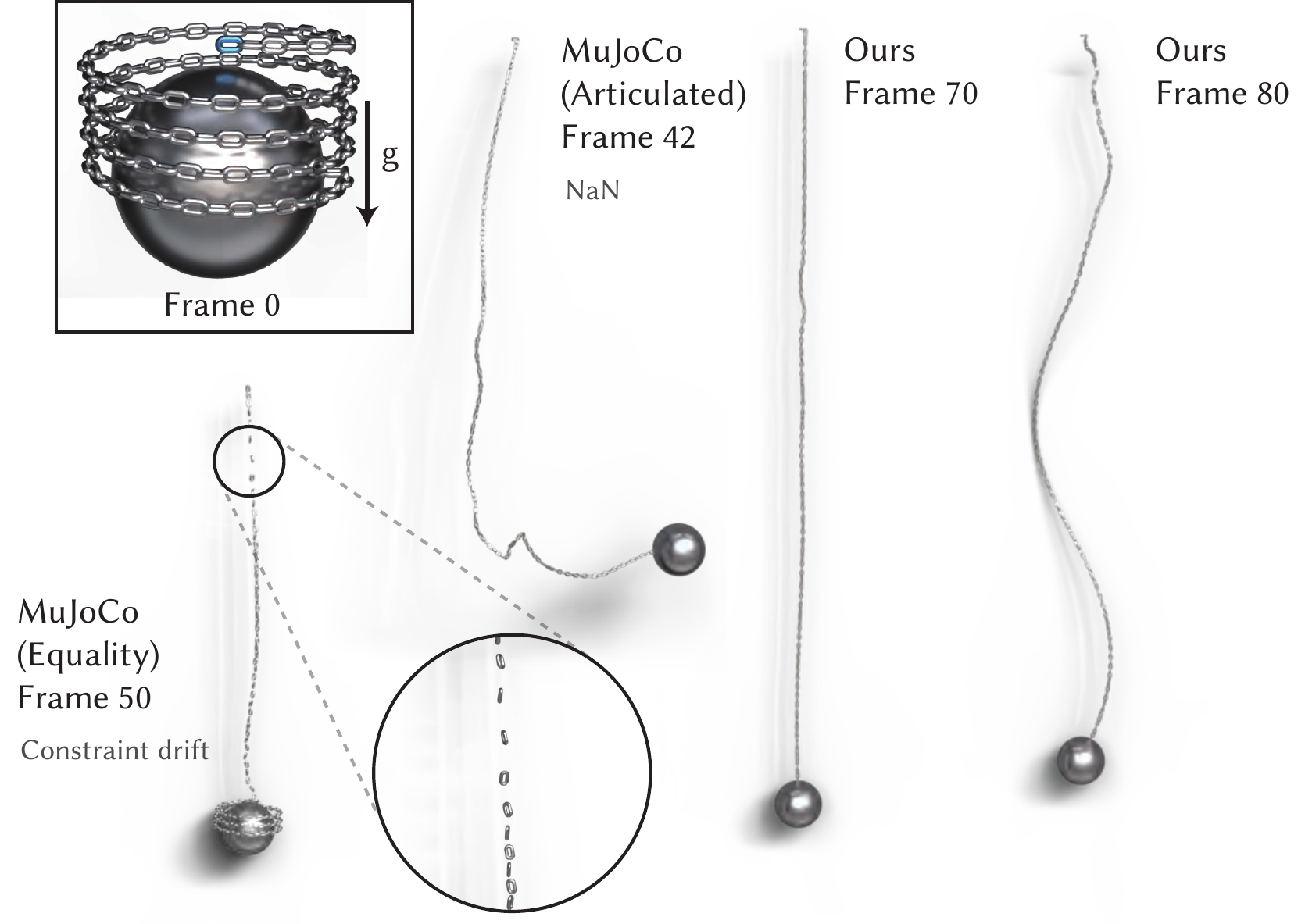}
\caption{\textbf{Chain with a heavy end mass.} 
We compare our method against two MuJoCo configurations (equality constraint and articulated joint). When $h = 10^{-3}~s$, MuJoCo's equality constraints exhibit joint drift and its articulated model fails to converge. Bullet and PhysX both crash in this experiment. Our method remains stable and naturally captures the elastic vibrations after the chain straightens even for $h = 10^{-2}~s$.
}
\Description{}
\label{fig:end mass}
\end{figure}
Fig.~\ref{fig:end mass} shows a long chain with a heavy end mass released from a coiled configuration. As the chain drops and straightens, it creates a massive impulsive load. We compare our results with two MuJoCo setups using either equality constraints or articulated joints. The time step size is $h=10^{-3}~s$ for all tests. Under sharp external impulses, MuJoCo's equality constraints show visible joint separation as seen in the zoomed view, while its articulated model becomes unstable and fails with a NaN error. In contrast, our method remains stable using only one iteration with all joint constraints exactly satisfied. The affine DOFs facilitate the conversion of impulse into elastic potential energy and eventually into kinetic energy. As a result, our method captures the natural vertical oscillations after the chain reaches full tension. We note that our method remains stable and produces the same result when $h = 10^{-2}~s$. In the example, PhysX and Bullet diverge. 

\begin{figure*}
\centering
\includegraphics[width=\linewidth]{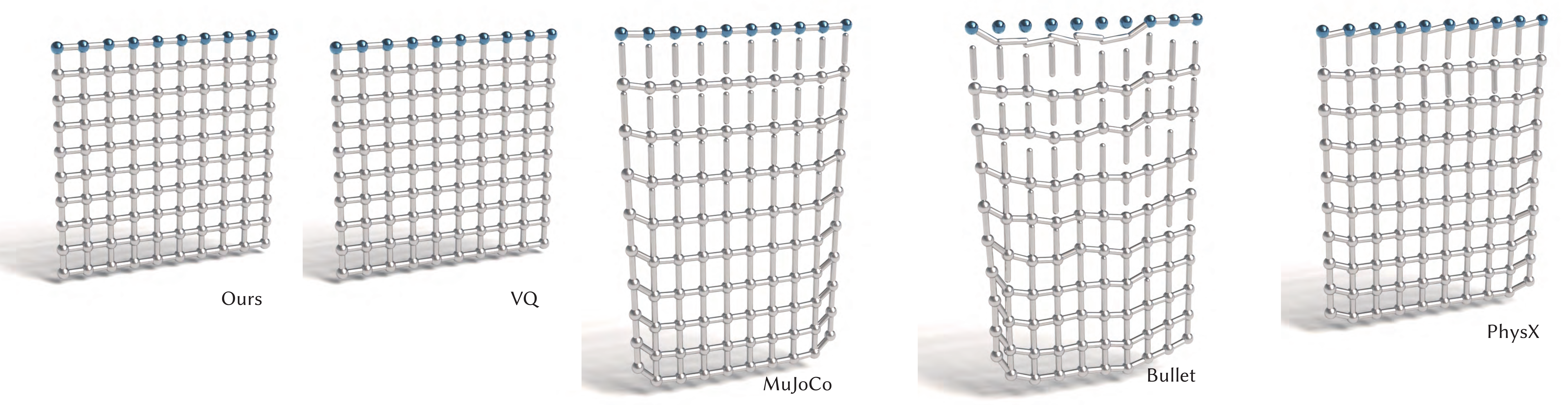}
\caption{\textbf{Ball-joint net I.} A $10\times 10$ net (280 links) is pinned along one boundary and sags under gravity with $h=1/30~s$. Our method uses a single iteration per step and produces a expected result. MuJoCo (30 iterations, equality constraint), Bullet (30 iterations), and PhysX (10 iterations) all fail to enforce the constraints and produce noticeable artifacts. VQ solves the global system using Newton's method. As a result, it also produces high-qualtiy result with all the constraints exactly satisfied. Nevertheless, it is much slower than our method. Our method uses less than  $1~ms$ for one step while VQ needs nearly $30~ms$.}
\Description{}
\label{fig:hanging_net}
\end{figure*}

\begin{figure}
\centering
\includegraphics[width=\linewidth]{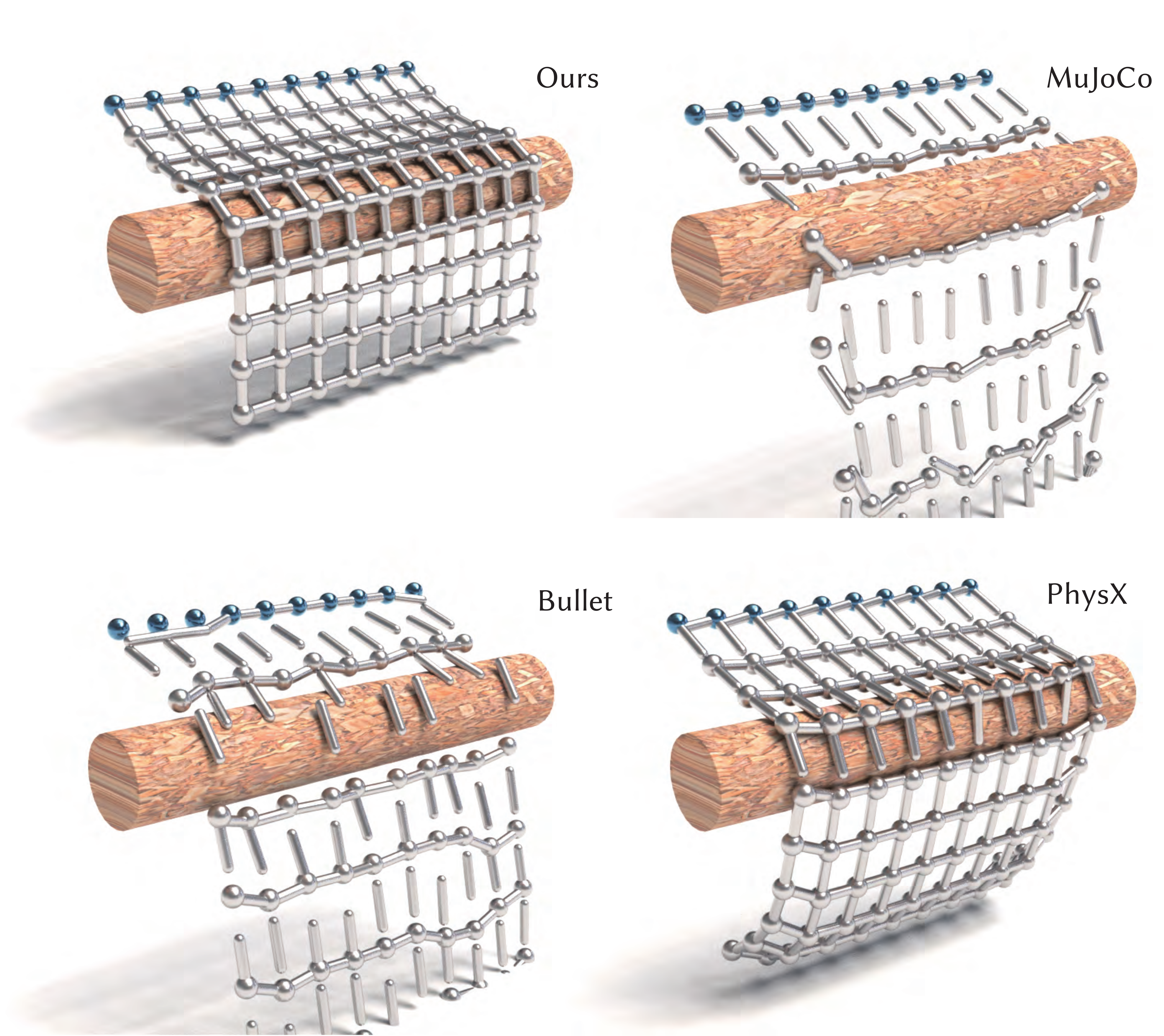}
\caption{\textbf{Ball-joint net II.} 
With an external collider, our method still performs with a single iteration per time step ($h = 1/30~s$. All other three methods fail to produce correct results.)
}
\Description{}
\label{fig:net_cylinder}
\end{figure}

\begin{figure}
\centering
\includegraphics[width=\linewidth]{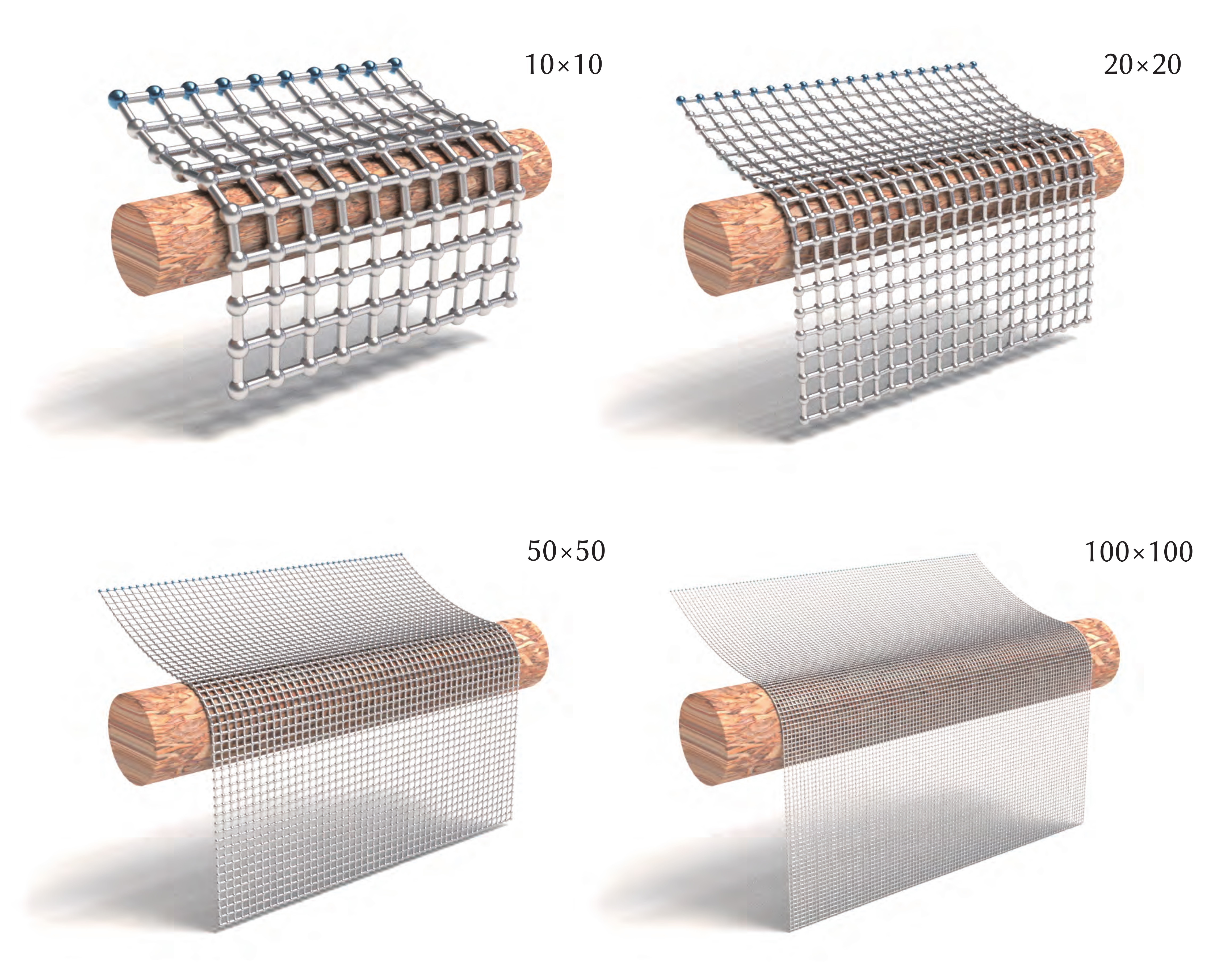}
\caption{\textbf{Ball-joint net III.} 
We increase the sizes of the net to $20\times 20$, $50\times 50$, and $100\times 100$ and simulate each case with a fixed time step size of $h=10^{-2}~s$ using one iteration. Our method remains stable while all the existing methods fail to handle such large-scale simulations.}
\Description{}
\label{fig:bignet_cylinder}
\end{figure}

Fig.~\ref{fig:hanging_net} shows a $10\times 10$ network of ball joints consisting of $280$ links. We use a relatively aggressive time step of $h=1/30~s$ to test robustness among various multibody frameworks. In this test, our method uses only one iteration per step, while PhysX uses $10$, and MuJoCo (equality constraint)/Bullet use $30$. Our method exactly enforces all joint constraints, preserving the net's geometry under gravity. On the other hand, all three other methods show visible gaps at the joints, indicating inaccurate constraint enforcement under these settings. VQ~\cite{VersatileQuaternion} solves the global system implicitly using Newton's method. It also yields high-quality results. Nevertheless, the matrix assembly and factorization make VQ much slower. In this example, our method uses less than $1~ms$ to finish one time step, while VQ needs $27~ms$. VQ quickly becomes prohibitive for large-scale instances such as the one shown in Figs.~\ref{fig:huge_chain} or \ref{fig:bignet_cylinder}.

In Fig.~\ref{fig:net_cylinder}, we put a cylinder collier under the ball-joint net to test contact-involved simulations. We use the same time step and iteration budgets as in Fig.~\ref{fig:hanging_net}. As the simulation becomes more complex, both MuJoCo and Bullet fail to converge, and the links become scattered. PhysX does not diverge, we can still see the joint gaps from the result. Our method remains stable with a single iteration --- the ball-joint net drapes on the collider naturally. We further test the scalability of our method. As shown in Fig.~\ref{fig:bignet_cylinder}, we increase the size of the ball-joint network to $20\times 20$, $50\times 50$, and $100\times 100$, respectively. We set $h=10^{-2}~s$ and use a single iteration per step for all cases. Despite the increased complexity, our method remains stable and produces coherent draping without any instabilities. All the other competitors fail to complete this example. 

\begin{figure*}
\centering
\includegraphics[width=\linewidth]{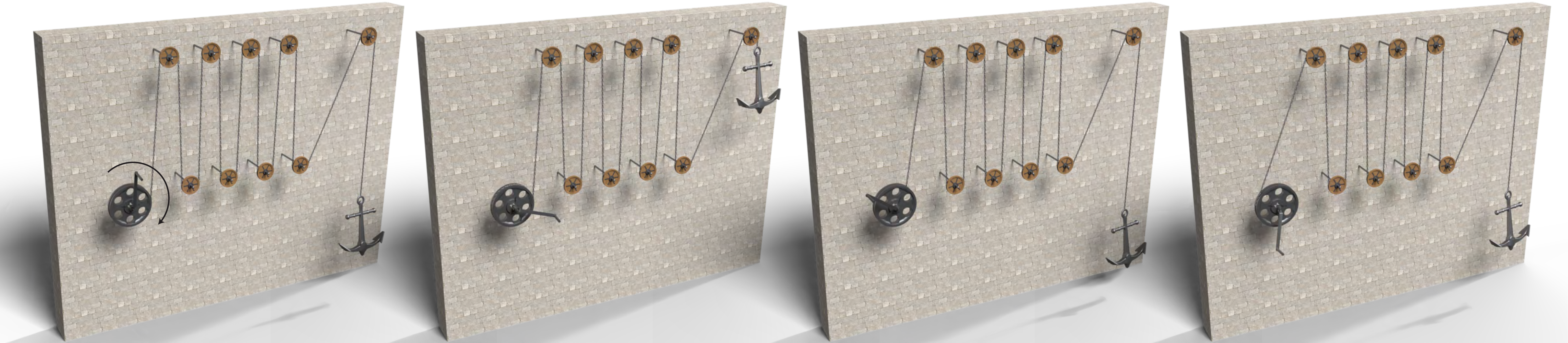}
\caption{\textbf{A pulley system.} We simulate a lifting-and-release cycle of a pulley system with a chain with $1.5$K joints. Our method stably handles the rapid tensioning as the heavy anchor rises and captures the inertial oscillations of the pulleys during the subsequent release. The time step size is $h=0.01~s$}
\Description{}
\label{fig:chain}
\end{figure*}
\subsection{Complex multi-ABD systems}
Our next set of experiments focuses on evaluating the proposed multi-ABD algorithm using the dual KKT solve under complex articulations and a large number of links. 

\begin{figure}
\centering
\includegraphics[width=\linewidth]{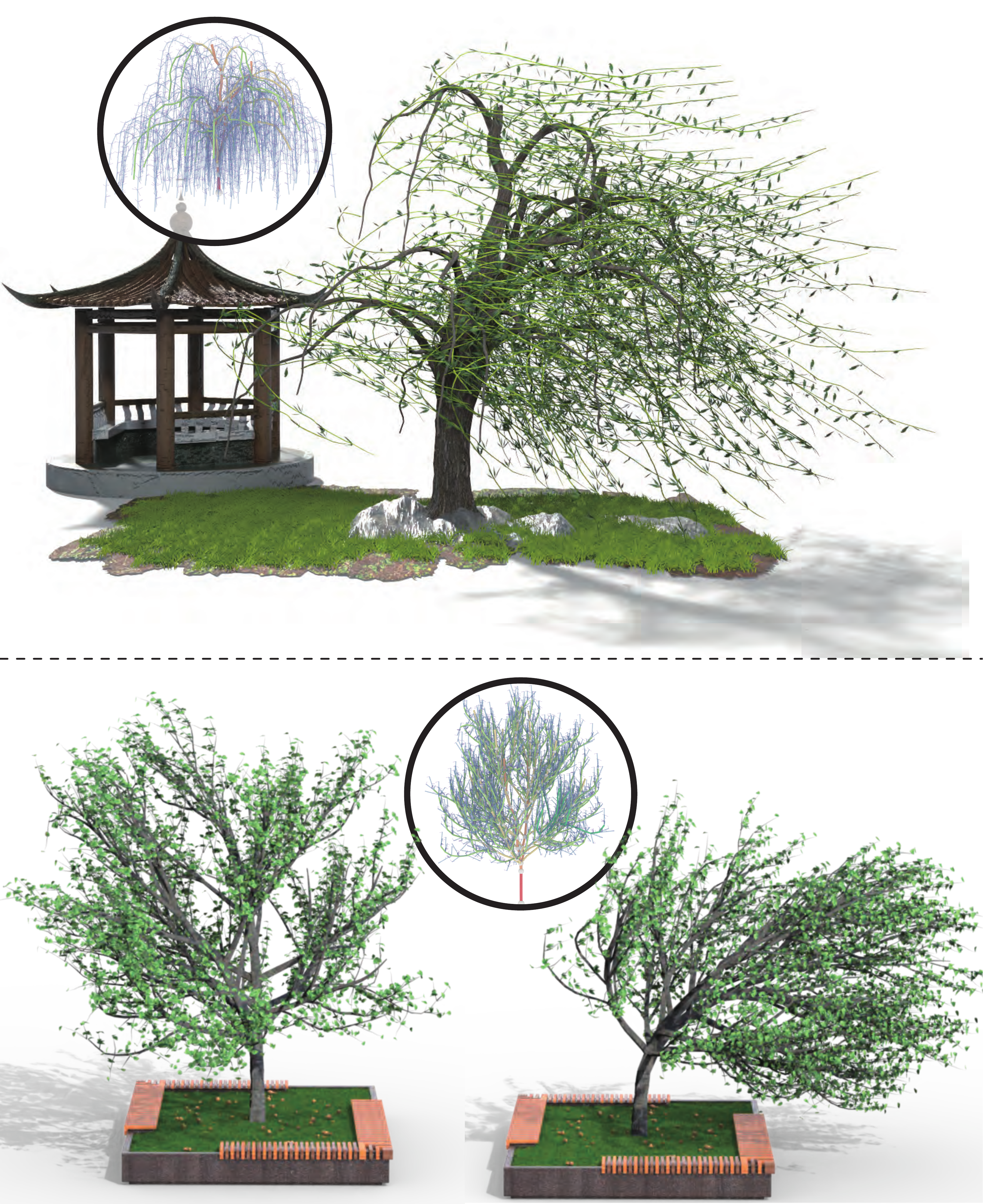}
\caption{\textbf{Tree as a multibody system.} We simulate a willow tree ($21$K links, top) and a pear tree ($29$K links, bottom) under winds. Our ABD-ABA framework handles these branched structures with a single iteration per step, requiring approximately $20~ms$ of computation time per frame on a single CPU thread. The resulting animation is high-quality and realistic. }
\Description{}
\label{fig:tree}
\end{figure}

\begin{figure}
\centering
\includegraphics[width=\linewidth]{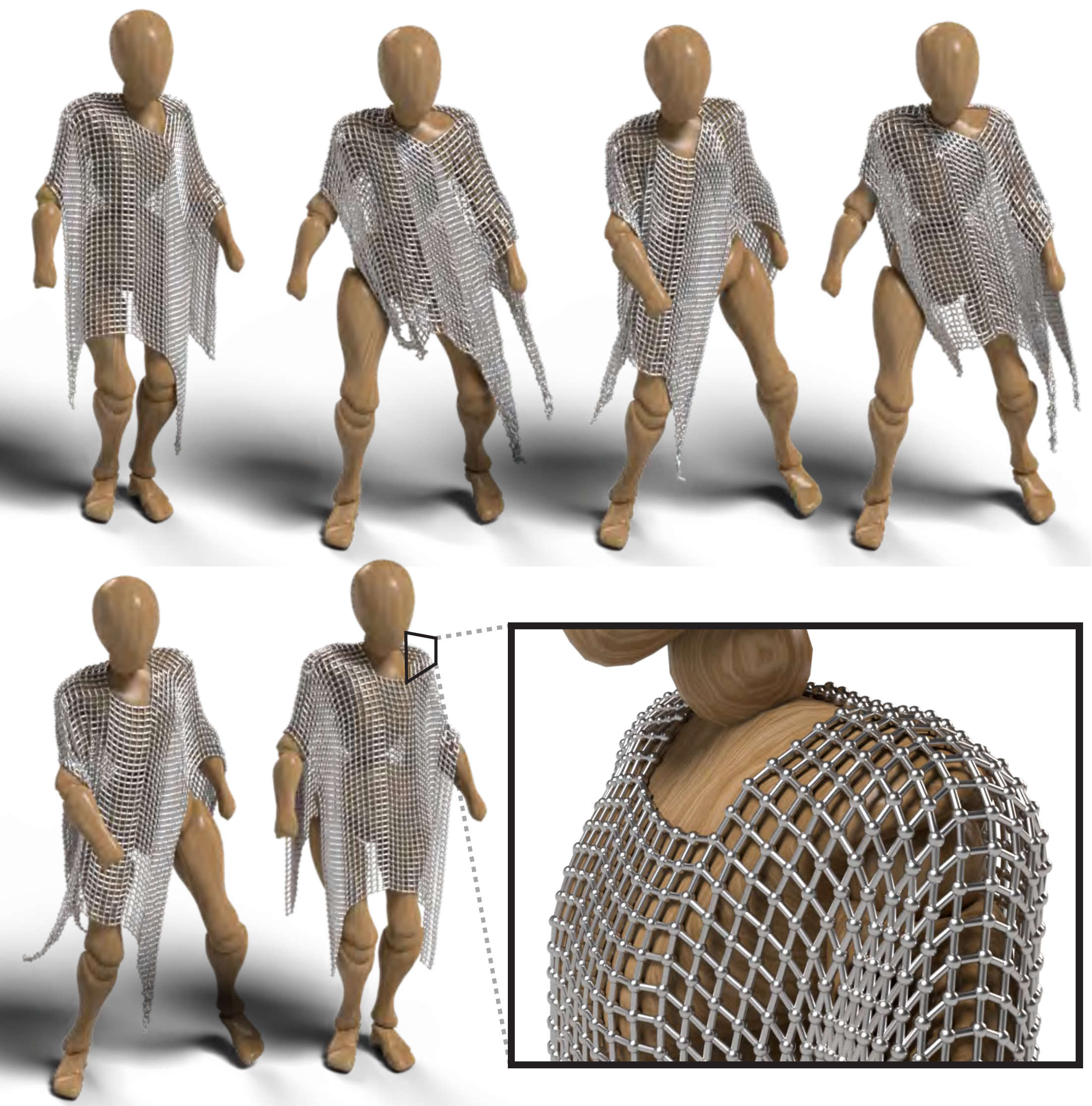}
\caption{\textbf{A multi-ABD cloak.}
A ``cloak'' is made of $11.7$K links connected with joints. It interacts with a moving avatar at $h=10^{-2}~s$. Our method captures the rich, garment-like dynamics of the structure while maintaining stable constraint enforcement and contact handling using only one iteration per step. MuJoCo, Bullet, and PhysX all fail in this example.}
\Description{}
\label{fig:net_cloth}
\end{figure}

\subsubsection{Joint chain.}
As discussed in Sec.~\ref{subsec:joint_chain}, our framework features a block-tridiagonal structure if an articulated system forms a chain of joints. This enables a linear-time solve only with one CPU thread. To validate this advantage, we showcase simulations of two pulley systems as shown in Figs.~\ref{fig:chain} and \ref{fig:huge_chain}. We set the time step size as $h = 10^{-2}~s$. Similar to Fig.~\ref{fig:bignet_cylinder}, MuJoCo, PhysX, and Bullet all fail in these examples (even with a highly conservative setting with $h = 10^{-4}~s$).

\begin{figure*}
\centering
\includegraphics[width=\linewidth]{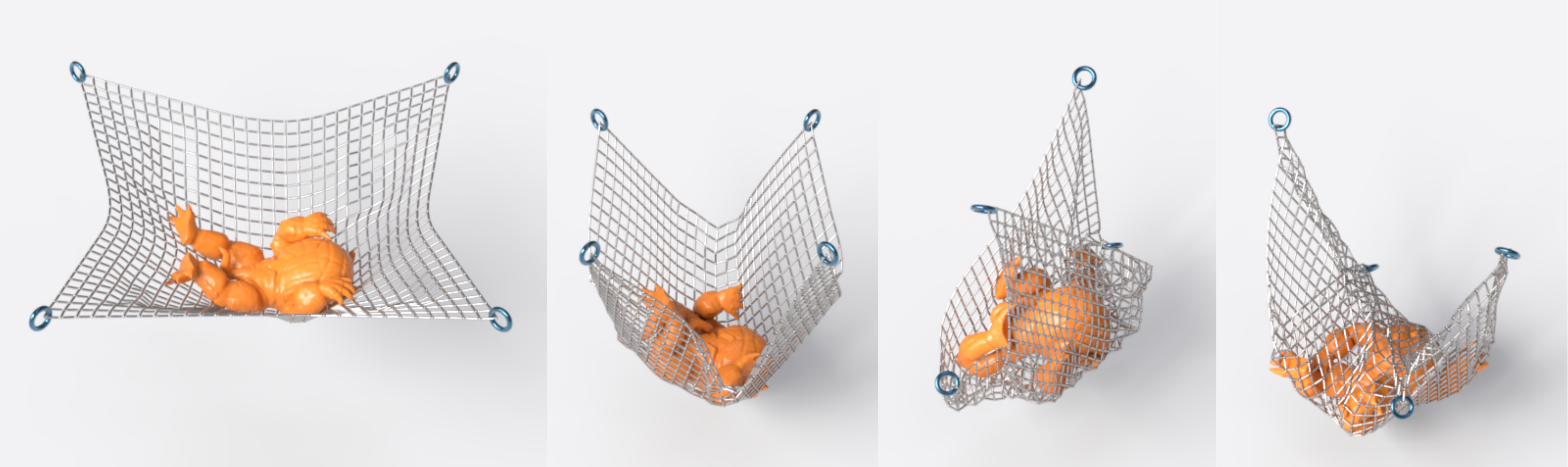}
\caption{\textbf{Armadillo.}
We simulate the interaction between a multi-ABD net and a Neo-Hookean Armadillo. Our method handles the strong coupling between articulated constraints and volumetric elasticity. With CP coordinate, ABD-based system can be seamlessly coupled with FEM models.}
\Description{}
\label{fig:armadillo}
\end{figure*}
\subsubsection{Joint graph.}
As the connectivity among joints becomes more complex, we need to solve the dual matrix of Eq.~\eqref{eq:dual}. If the system size is moderate, a direct solver could be used. Otherwise, we can use a multi-directional GS procedure to break to solve along multiple joint threads.  
Fig.~\ref{fig:chain} shows a lifting-and-release cycle of a pulley system consisting of $1.5$K joints. The major portion of DOFs is on the long chain linked to a heavy anchor. During loading, the chain tightens and the anchor rises as tension moves through the pulleys. Once the drive is released, the anchor falls and the chain slides back. A brief back-and-forth motion occurs during unloading, caused by the redistribution of tension and the pulleys' inertia. Fig.~\ref{fig:huge_chain} shows a stress test of a similar pulley setup, scaled up to $1,076,748$ links. Our method remains stable with $h = 10^{-2}~s$ and only needs one iteration for each step. Meanwhile, all the joint constraints are exactly enforced. Running at a single CPU thread, our method only takes $904~ms$ on average to simulate one step.

\subsubsection{Joint hierarchy.}
Our ABD-ABA framework generalizes joint chain systems to multi-level topologies, allowing it to handle large-scale, complex joint hierarchies (Sec.~\ref{subsec:tree}). Fig.~\ref{fig:tree} showcases this with a willow tree ($21$K links) and a pear tree ($29$K links). To build these models, we extract skeletal graphs from the tree geometries (as highlighted in the figure) and organize the links by branching depth for efficient traversal. Each joint includes penalty energies to model real-world branch compliance. A prescribed wind field applies aerodynamic loads, which are accumulated as external wrenches and mapped to generalized forces in ABD coordinates. Our solver advances the state at $h = 10^{-2}$ s using only a single iteration per step. The simulation takes approximately $20~ms$ per step for these models, which provides an alternative approach to real-time simulation of complex plants.

\begin{figure}
\centering
\includegraphics[width=\linewidth]{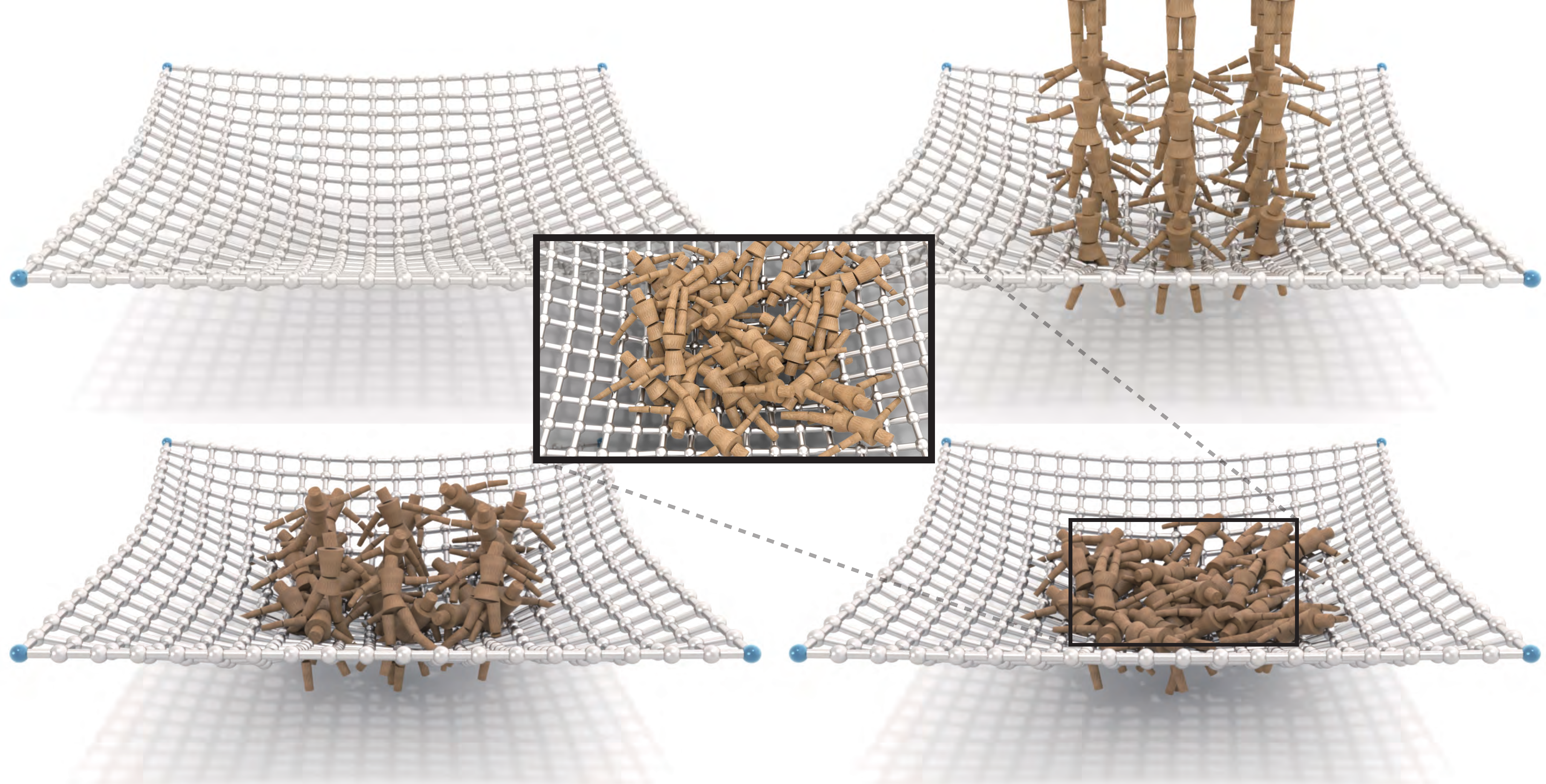}
\caption{\textbf{Ragdolls.} We drop $27$ ragdolls onto a stiff (with Young's modulus being $10^6$) joint net. Our method gives stable results under high-speed collisions. In this experiment, we have time step size $h=1/200~s$, and our simulation uses one iteration for each step.}
\Description{}
\label{fig:ragdolls}
\end{figure}

A representative example is shown in Fig.~\ref{fig:net_cloth}, where a cloak-like jointed net interacts with a moving avatar. The cloak consists of $11,750$ links that slide frictionlessly over the body. As in previous examples, we set $h=10^{-2}~s$ and use only one iteration per step. Despite being a collection of rigid links and joints, the entire structure behaves like a flexible garment, producing rich dynamic effects as it follows the body's movement. Notably, none of the other simulators i.e., MuJoCo, Bullet, or PhysX manage to run this example, whereas our method remains stable throughout the sequence. Fig.~\ref{fig:ragdolls} gives another example, where $27$ ragdolls fall onto a corner-pinned ball-joint net ($1,160$ links). Each ragdoll is an $11$-link articulated body with joint penalties for bending and twisting. We use a reduced Young's modulus of $10^6$ to allow small deflections of the network. The experiment involves dense impacts and changing contact topologies between the ragdolls and the net, as well as among the ragdolls themselves. We use a smaller time step size of $h=1/200~s$ to facilitate the collision detection of fast-movement ragdolls. Our method uses $33~ms$ to process one step.

\subsection{More versatile results}
ABD formulation shares the same underlying framework as FEM (finite element method). The control tetrahedron of an affine body essentially makes it a single-element model. Therefore, ABD-based multibody systems naturally integrate with deformable body simulations. Fig.~\ref{fig:armadillo} demonstrates the coupling between a multi-ABD net and a Neo-Hookean Armadillo. In this example, we drop the Armadillo onto the net and then twist the boundary anchors to induce shear and torsion. This example highlights the stability of our solver under strong coupling between articulated constraints and volumetric elasticity.

\begin{figure*}
\centering
\includegraphics[width=\linewidth]{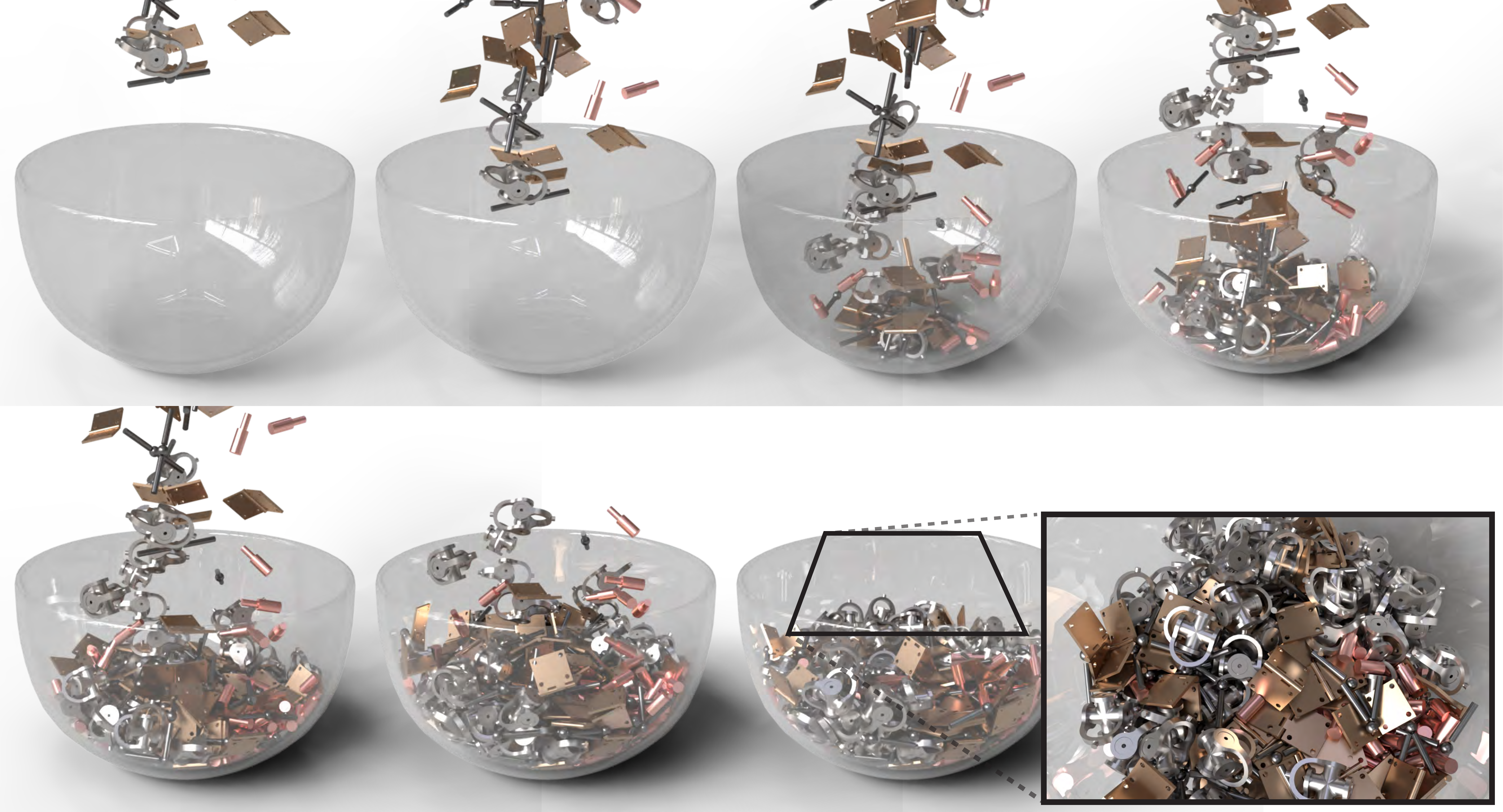}
\caption{\textbf{Falling joints.} We drop $720$ articulated bodies connected by different joint types, including $180$ ball joints, $180$ universal joints, $180$ hinge joints, and $180$ prismatic joints. Our solver robustly maintains all joint configurations under dense stacking and repeated impacts. 
}
\Description{}
\label{fig:falling_joints}
\end{figure*}

Fig.~\ref{fig:falling_joints} shows snapshots of another collision-heavy scene. In this experiment,  we drop $720$ jointed pairs into a bowl container, including $180$ ball joints, $180$ universal joints, $180$ hinge joints, and $180$ prismatic joints. All the constraints are exactly satisfied during the simulation, and we use the implicit penalty method to handle collisions among articulated bodies and collisions with the collider. 

\begin{figure*}
\centering
\includegraphics[width=\linewidth]{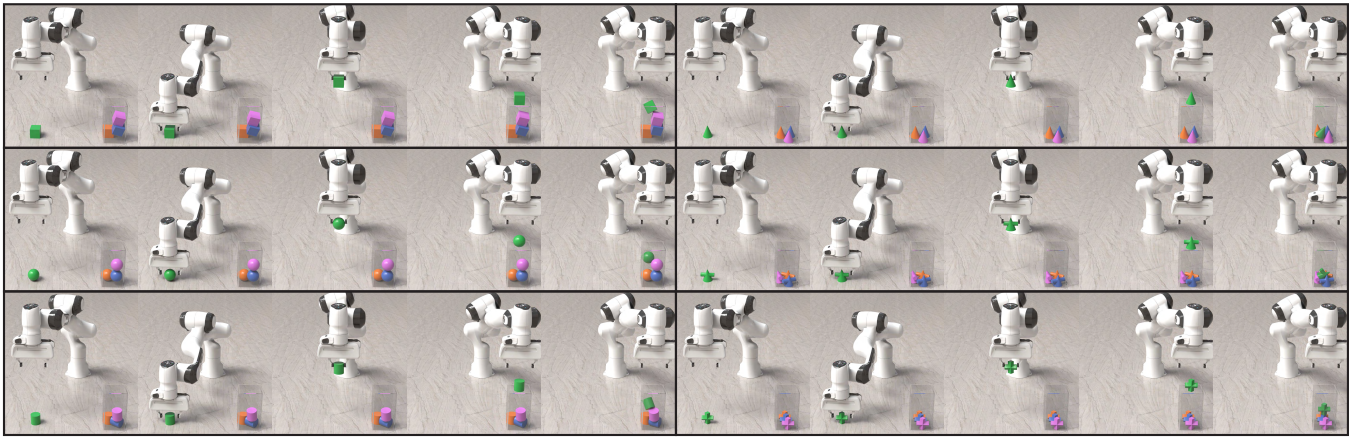}
\caption{\textbf{Multi-body ABD for Embodied AI.}
We integrate our ABD solver into a standard robotics pipeline to perform pick-and-place tasks. Even when allocated limited computational resources, specifically a single iteration per step, our method maintains the joint integrity and contact stability required for training and testing Embodied AI agents. The solver robustly handles the transition from planned motion to contact-rich interaction as objects are grasped, transported, and stacked.
}
\Description{}
\label{fig:pick}
\end{figure*}

A motivation for our ABD-based multibody solver is to provide an efficient and robust algorithm for Embodied AI training and testing. In many large-scale or real-time applications, each simulation scene is allocated with limited computational resources. Our framework is designed to maintain physical stability and constraint integrity even under these constraints, using minimal solver iterations and large time steps. This robustness allows for the reliable evaluation of agents across diverse, contact-rich environments without the numerical failures that often plague traditional simulators in resource-constrained scenarios. To this end, Fig.~\ref{fig:pick} showcases this efficiency in a robotics context. We integrate our ABD framework with inverse kinematics (IK) and motion planning to drive a Franka Panda manipulator through pick-and-place tasks. Our solver handles joint actuation and contact interactions between the gripper and objects using a single iteration per step. The results show that the arm reliably executes grasping and transport while objects remain stable during stacking and settling. 

\begin{figure*}
\centering
\includegraphics[width=\linewidth]{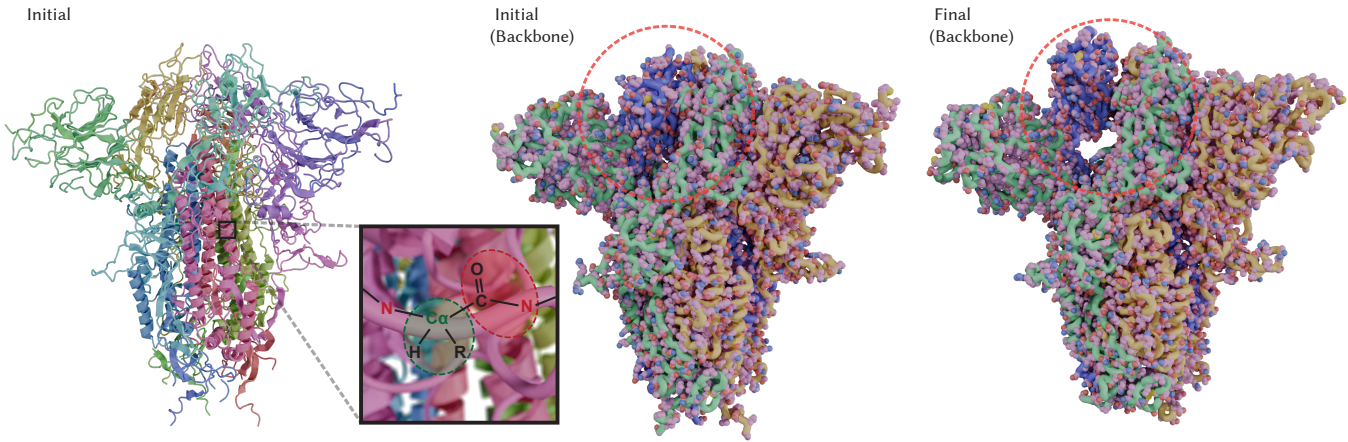}
\caption{\textbf{Protein unfolding.} A polypeptide backbone is modeled as an articulated chain of rigid links connected by ball joints parameterized by the dihedral angles $\phi$ and $\psi$, which yields a rigid-link abstraction of backbone motion. Given a sparse time sequence of marker observations $\{\bm y_t\}$ (e.g., $\mathrm{C}_\alpha$ positions), we solve an IK problem to recover generalized coordinates $\bm q_t$ at the observed frames, and we then roll out the ABD dynamics between these IK keyframes to obtain a temporally continuous backbone motion.
Left: rendered initial protein structure with a zoom-in illustrating the locally rigid backbone geometry used to define each link.
Middle and right: backbone representations at the initial and final frames of the reconstructed sequence. A region that undergoes a large conformational change between the initial and final states is highlighted.}
\Description{}
\label{fig:protein}
\end{figure*}
Fig.~\ref{fig:protein} demonstrates a cross-disciplinary application of our framework in structural biology. A polypeptide backbone can be modeled as an articulated chain because its flexibility is primarily determined by a subset of rotatable bonds. We represent the backbone as a sequence of rigid units connected by rotatable dihedrals $\phi$ and $\psi$. 
In this example, the original protein dynamics sequence is a 40\,ns trajectory of the SARS-CoV-2 Spike glycoprotein (PDB: 6VXX), used to generate temporally sparse observations; for biological context on SARS-CoV-2 recognition by full-length human ACE2, see ~\citet{Yan2020ACE2,Walls2020Spike}.
To reconstruct the motion of a viral protein undergoing a large-scale conformational change, we solve for IK keyframes from sparse observations and roll out multi-ABD dynamics to recover a continuous trajectory. This procedure produces a smooth animation that fills the gaps between temporally sparse markers. The recovered motion maintains the folded topology and structural coherence without any discontinuities or numerical fragmentation. This result illustrates that the proposed ABD-based multibody framework is a versatile tool for simulating complex articulated systems across different scientific domains, including the study of large-scale structural transitions in biological macromolecules.

\section{Conclusion \& Limitations}\label{sec:con}
This paper presents an efficient framework for simulating large-scale articulated systems. By separating geometric nonlinearity with a co-rotational formulation, we turn the ABD problem into a constant-matrix system, which can be pre-factorized even with implicit integration. Our analysis of joint constraints in the affine coordinate space provides both linear and nonlinear constraint models, which allows the simulation to capture complex mechanical interactions accurately. To handle different mechanical structures, we developed a collection of specialized solvers, including a block-tridiagonal solver, a generalized Featherstone algorithm, a low-dimensional approximation for kinetic loops, and a bidirectional Gauss-Seidel optimizer for dense joint networks. By projecting the problem into a minimal-DOF dual space, it is possible to solve the KKT on a single-threaded CPU at a scale not previously achieved. We believe our method offers a novel and reliable solution for many demanding tasks in embodied AI, mechanical design, and interactive graphics. 

Our method also has limitations. While all the equality constraints induced by joints are exactly and robustly processed with KKT, inequality constraints remain a challenge. If two bodies are about to colliding with each other, it is possible to employ collision barriers~\cite{IPC} to resolve the collision. However, doing so puts nonlinear off-diagonal entries in the primal matrix, and our pre-factorized per-body solve is less helpful. In order to fully exploit the co-rotated ABD model, we plan to generalize the proposed primal-dual algorithm for inequality constraints. This strategy will need to pivot active primal DOFs, and the key problem is how to fast identify the right set of DOFs to be activated (i.e., converting to equality constraints). Our method solves the dual matrix using Newton's method on the CPU. It is also possible to use GPU-based solvers to further accelerate the computation. This is particularly useful for large-scale molecular simulations such as the protein example in Fig.~\ref{fig:protein}.


\bibliographystyle{ACM-Reference-Format}
\bibliography{ref}

@article{ABD,
author = {Lan, Lei and Kaufman, Danny M. and Li, Minchen and Jiang, Chenfanfu and Yang, Yin},
title = {Affine body dynamics: fast, stable and intersection-free simulation of stiff materials},
year = {2022},
issue_date = {July 2022},
publisher = {Association for Computing Machinery},
address = {New York, NY, USA},
volume = {41},
number = {4},
issn = {0730-0301},
url = {https://doi.org/10.1145/3528223.3530064},
doi = {10.1145/3528223.3530064},
journal = {ACM Trans. Graph.},
month = jul,
articleno = {67},
numpages = {14}
}

@article{IPC,
author = {Li, Minchen and Ferguson, Zachary and Schneider, Teseo and Langlois, Timothy and Zorin, Denis and Panozzo, Daniele and Jiang, Chenfanfu and Kaufman, Danny M.},
title = {Incremental potential contact: intersection-and inversion-free, large-deformation dynamics},
year = {2020},
issue_date = {August 2020},
publisher = {Association for Computing Machinery},
address = {New York, NY, USA},
volume = {39},
number = {4},
issn = {0730-0301},
url = {https://doi.org/10.1145/3386569.3392425},
doi = {10.1145/3386569.3392425},
journal = {ACM Trans. Graph.},
month = aug,
articleno = {49},
numpages = {20}
}

@article{UnifiedNewtonBarrier,
author = {Chen, Yunuo and Li, Minchen and Lan, Lei and Su, Hao and Yang, Yin and Jiang, Chenfanfu},
title = {A unified newton barrier method for multibody dynamics},
year = {2022},
issue_date = {July 2022},
publisher = {Association for Computing Machinery},
address = {New York, NY, USA},
volume = {41},
number = {4},
issn = {0730-0301},
url = {https://doi.org/10.1145/3528223.3530076},
doi = {10.1145/3528223.3530076},
journal = {ACM Trans. Graph.},
month = jul,
articleno = {66},
numpages = {14}
}

@article{Wang2019RedMax,
author = {Wang, Ying and Weidner, Nicholas J. and Baxter, Margaret A. and Hwang, Yura and Kaufman, Danny M. and Sueda, Shinjiro},
title = {RedMax: efficient \& flexible approach for articulated dynamics},
year = {2019},
issue_date = {August 2019},
publisher = {Association for Computing Machinery},
address = {New York, NY, USA},
volume = {38},
number = {4},
issn = {0730-0301},
url = {https://doi.org/10.1145/3306346.3322952},
doi = {10.1145/3306346.3322952},
journal = {ACM Trans. Graph.},
month = jul,
articleno = {104},
numpages = {10}
}

@article{Geilinger2020ADD,
author = {Geilinger, Moritz and Hahn, David and Zehnder, Jonas and B\"{a}cher, Moritz and Thomaszewski, Bernhard and Coros, Stelian},
title = {ADD: analytically differentiable dynamics for multi-body systems with frictional contact},
year = {2020},
issue_date = {December 2020},
publisher = {Association for Computing Machinery},
address = {New York, NY, USA},
volume = {39},
number = {6},
issn = {0730-0301},
url = {https://doi.org/10.1145/3414685.3417766},
doi = {10.1145/3414685.3417766},
journal = {ACM Trans. Graph.},
month = nov,
articleno = {190},
numpages = {15}
}

@article{Macklin2019NonSmoothNewton,
author = {Macklin, Miles and Erleben, Kenny and M\"{u}ller, Matthias and Chentanez, Nuttapong and Jeschke, Stefan and Makoviychuk, Viktor},
title = {Non-smooth Newton Methods for Deformable Multi-body Dynamics},
year = {2019},
issue_date = {October 2019},
publisher = {Association for Computing Machinery},
address = {New York, NY, USA},
volume = {38},
number = {5},
issn = {0730-0301},
url = {https://doi.org/10.1145/3338695},
doi = {10.1145/3338695},
journal = {ACM Trans. Graph.},
month = oct,
articleno = {140},
numpages = {20}
}

@article{PBD,
author = {M\"{u}ller, Matthias and Heidelberger, Bruno and Hennix, Marcus and Ratcliff, John},
title = {Position based dynamics},
year = {2007},
issue_date = {April, 2007},
publisher = {Academic Press, Inc.},
address = {USA},
volume = {18},
number = {2},
issn = {1047-3203},
url = {https://doi.org/10.1016/j.jvcir.2007.01.005},
doi = {10.1016/j.jvcir.2007.01.005},
journal = {J. Vis. Comun. Image Represent.},
month = apr,
pages = {109–118},
numpages = {10}
}

@article{Baraff1989,
author = {Baraff, D.},
title = {Analytical methods for dynamic simulation of non-penetrating rigid bodies},
year = {1989},
issue_date = {July 1989},
publisher = {Association for Computing Machinery},
address = {New York, NY, USA},
volume = {23},
number = {3},
issn = {0097-8930},
url = {https://doi.org/10.1145/74334.74356},
doi = {10.1145/74334.74356},
journal = {SIGGRAPH Comput. Graph.},
month = jul,
pages = {223–232},
numpages = {10}
}

@InProceedings{Moreau1985,
author="Moreau, J. J.",
editor="Del Piero, Gianpietro
and Maceri, Franco",
title="Standard Inelastic Shocks and the Dynamics of Unilateral Constraints",
booktitle="Unilateral Problems in Structural Analysis",
year="1985",
publisher="Springer Vienna",
address="Vienna",
pages="173--221",
isbn="978-3-7091-2632-5"
}

@article{Stewart2000,
author = {Stewart, David E.},
title = {Rigid-Body Dynamics with Friction and Impact},
journal = {SIAM Review},
volume = {42},
number = {1},
pages = {3-39},
year = {2000},
doi = {10.1137/S0036144599360110},

URL = { 
        https://doi.org/10.1137/S0036144599360110
},
eprint = { 
        https://doi.org/10.1137/S0036144599360110
}
}

@article{Smith2012,
author = {Smith, Breannan and Kaufman, Danny M. and Vouga, Etienne and Tamstorf, Rasmus and Grinspun, Eitan},
title = {Reflections on simultaneous impact},
year = {2012},
issue_date = {July 2012},
publisher = {Association for Computing Machinery},
address = {New York, NY, USA},
volume = {31},
number = {4},
issn = {0730-0301},
url = {https://doi.org/10.1145/2185520.2185602},
doi = {10.1145/2185520.2185602},
journal = {ACM Trans. Graph.},
month = jul,
articleno = {106},
numpages = {12}
}

@techreport{Mirtich1994,
    Author= {Mirtich, Brian and Canny, John F.},
    Title= {Impulse-based Dynamic Simulation},
    Year= {1994},
    Month= {Jun},
    Url= {http://www2.eecs.berkeley.edu/Pubs/TechRpts/1994/5494.html},
    Number= {UCB/CSD-94-815},
}

@ARTICLE{Weinstein2006,
  author={Weinstein, R. and Teran, J. and Fedkiw, R.},
  journal={IEEE Transactions on Visualization and Computer Graphics}, 
  title={Dynamic simulation of articulated rigid bodies with contact and collision}, 
  year={2006},
  volume={12},
  number={3},
  pages={365-374},
  doi={10.1109/TVCG.2006.48}}

@inproceedings{Redon2005,
author = {Redon, Stephane and Galoppo, Nico and Lin, Ming C.},
title = {Adaptive dynamics of articulated bodies},
year = {2005},
isbn = {9781450378253},
publisher = {Association for Computing Machinery},
address = {New York, NY, USA},
url = {https://doi.org/10.1145/1186822.1073294},
doi = {10.1145/1186822.1073294},
booktitle = {ACM SIGGRAPH 2005 Papers},
pages = {936–945},
numpages = {10},
location = {Los Angeles, California},
series = {SIGGRAPH '05}
}

@article{Sueda2011,
author = {Sueda, Shinjiro and Jones, Garrett L. and Levin, David I. W. and Pai, Dinesh K.},
title = {Large-scale dynamic simulation of highly constrained strands},
year = {2011},
issue_date = {July 2011},
publisher = {Association for Computing Machinery},
address = {New York, NY, USA},
volume = {30},
number = {4},
issn = {0730-0301},
url = {https://doi.org/10.1145/2010324.1964934},
doi = {10.1145/2010324.1964934},
journal = {ACM Trans. Graph.},
month = jul,
articleno = {39},
numpages = {10}
}

@article{Tournier2015,
author = {Tournier, Maxime and Nesme, Matthieu and Gilles, Benjamin and Faure, Fran\c{c}ois},
title = {Stable constrained dynamics},
year = {2015},
issue_date = {August 2015},
publisher = {Association for Computing Machinery},
address = {New York, NY, USA},
volume = {34},
number = {4},
issn = {0730-0301},
url = {https://doi.org/10.1145/2766969},
doi = {10.1145/2766969},
journal = {ACM Trans. Graph.},
month = jul,
articleno = {132},
numpages = {10}
}

@article{Andrews2017,
author = {Andrews, Sheldon and Teichmann, Marek and Kry, Paul G.},
title = {Geometric Stiffness for Real-time Constrained Multibody Dynamics},
journal = {Computer Graphics Forum},
volume = {36},
number = {2},
pages = {235-246},
doi = {https://doi.org/10.1111/cgf.13122},
url = {https://onlinelibrary.wiley.com/doi/abs/10.1111/cgf.13122},
eprint = {https://onlinelibrary.wiley.com/doi/pdf/10.1111/cgf.13122},
year = {2017}
}

@inproceedings{Enzenhofer2019,
author = {Enzenh\"{o}fer, Andreas and Lefebvre, Nicolas and Andrews, Sheldon},
title = {Efficient block pivoting for multibody simulations with contact},
year = {2019},
isbn = {9781450363105},
publisher = {Association for Computing Machinery},
address = {New York, NY, USA},
url = {https://doi.org/10.1145/3306131.3317019},
doi = {10.1145/3306131.3317019},
booktitle = {Proceedings of the ACM SIGGRAPH Symposium on Interactive 3D Graphics and Games},
articleno = {2},
numpages = {9},
location = {Montreal, Quebec, Canada},
series = {I3D '19}
}

@article{Peiret2019,
author = {Peiret, Albert and Andrews, Sheldon and K\"{o}vecses, J\'{o}zsef and Kry, Paul G. and Teichmann, Marek},
title = {Schur Complement-based Substructuring of Stiff Multibody Systems with Contact},
year = {2019},
issue_date = {October 2019},
publisher = {Association for Computing Machinery},
address = {New York, NY, USA},
volume = {38},
number = {5},
issn = {0730-0301},
url = {https://doi.org/10.1145/3355621},
doi = {10.1145/3355621},
journal = {ACM Trans. Graph.},
month = oct,
articleno = {150},
numpages = {17}
}

@article{Werling2021,
  author       = {Keenon Werling and
                  Dalton Omens and
                  Jeongseok Lee and
                  Ioannis Exarchos and
                  C. Karen Liu},
  title        = {Fast and Feature-Complete Differentiable Physics for Articulated Rigid
                  Bodies with Contact},
  journal      = {CoRR},
  volume       = {abs/2103.16021},
  year         = {2021},
  url          = {https://arxiv.org/abs/2103.16021},
  eprinttype    = {arXiv},
  eprint       = {2103.16021},
  bibsource    = {dblp computer science bibliography, https://dblp.org}
}

@article{VersatileQuaternion,
author = {Maloisel, Guirec and Grandia, Ruben and Schumacher, Christian and Knoop, Espen and B\"{a}cher, Moritz},
title = {A Versatile Quaternion-Based Constrained Rigid Body Dynamics},
year = {2025},
issue_date = {August 2025},
publisher = {Association for Computing Machinery},
address = {New York, NY, USA},
volume = {44},
number = {4},
issn = {0730-0301},
url = {https://doi.org/10.1145/3730872},
doi = {10.1145/3730872},
journal = {ACM Trans. Graph.},
month = jul,
articleno = {156},
numpages = {17}
}

@inproceedings{Shinar2008,
author = {Shinar, Tamar and Schroeder, Craig and Fedkiw, Ronald},
title = {Two-way coupling of rigid and deformable bodies},
year = {2008},
isbn = {9783905674101},
publisher = {Eurographics Association},
address = {Goslar, DEU},
booktitle = {Proceedings of the 2008 ACM SIGGRAPH/Eurographics Symposium on Computer Animation},
pages = {95–103},
numpages = {9},
location = {Dublin, Ireland},
series = {SCA '08}
}

@inproceedings{Kaufman2008,
author = {Kaufman, Danny M. and Sueda, Shinjiro and James, Doug L. and Pai, Dinesh K.},
title = {Staggered projections for frictional contact in multibody systems},
year = {2008},
isbn = {9781450318310},
publisher = {Association for Computing Machinery},
address = {New York, NY, USA},
url = {https://doi.org/10.1145/1457515.1409117},
doi = {10.1145/1457515.1409117},
booktitle = {ACM SIGGRAPH Asia 2008 Papers},
articleno = {164},
numpages = {11},
location = {Singapore},
series = {SIGGRAPH Asia '08}
}

@inproceedings{JainLiu2011,
author = {Jain, Sumit and Liu, C. Karen},
title = {Controlling physics-based characters using soft contacts},
year = {2011},
isbn = {9781450308076},
publisher = {Association for Computing Machinery},
address = {New York, NY, USA},
url = {https://doi.org/10.1145/2024156.2024197},
doi = {10.1145/2024156.2024197},
booktitle = {Proceedings of the 2011 SIGGRAPH Asia Conference},
articleno = {163},
numpages = {10},
location = {Hong Kong, China},
series = {SA '11}
}

@inproceedings{BaiLiu2014,
author = {Bai, Yunfei and Liu, C. Karen},
title = {Coupling cloth and rigid bodies for dexterous manipulation},
year = {2014},
isbn = {9781450326230},
publisher = {Association for Computing Machinery},
address = {New York, NY, USA},
url = {https://doi.org/10.1145/2668064.2668066},
doi = {10.1145/2668064.2668066},
booktitle = {Proceedings of the 7th International Conference on Motion in Games},
pages = {139–145},
numpages = {7},
location = {Playa Vista, California},
series = {MIG '14}
}

@article{Deul2016,
author = {Deul, Crispin and Charrier, Patrick and Bender, Jan},
title = {Position-based rigid-body dynamics},
year = {2016},
issue_date = {March 2016},
publisher = {John Wiley and Sons Ltd.},
address = {GBR},
volume = {27},
number = {2},
issn = {1546-4261},
url = {https://doi.org/10.1002/cav.1614},
doi = {10.1002/cav.1614},
journal = {Comput. Animat. Virtual Worlds},
month = mar,
pages = {103–112},
numpages = {10}
}

@inproceedings{Muller2020,
author = {M\"{u}ller, Matthias and Macklin, Miles and Chentanez, Nuttapong and Jeschke, Stefan and Kim, Tae-Yong},
title = {Detailed rigid body simulation with extended position based dynamics},
year = {2020},
publisher = {Eurographics Association},
address = {Goslar, DEU},
url = {https://doi.org/10.1111/cgf.14105},
doi = {10.1111/cgf.14105},
booktitle = {Proceedings of the ACM SIGGRAPH/Eurographics Symposium on Computer Animation},
articleno = {10},
numpages = {12},
location = {Virtual Event, Canada},
series = {SCA '20}
}

@article{CIPC,
author = {Li, Minchen and Kaufman, Danny M. and Jiang, Chenfanfu},
title = {Codimensional incremental potential contact},
year = {2021},
issue_date = {August 2021},
publisher = {Association for Computing Machinery},
address = {New York, NY, USA},
volume = {40},
number = {4},
issn = {0730-0301},
url = {https://doi.org/10.1145/3450626.3459767},
doi = {10.1145/3450626.3459767},
journal = {ACM Trans. Graph.},
month = jul,
articleno = {170},
numpages = {24}
}

@article{rigidIPC,
author = {Ferguson, Zachary and Li, Minchen and Schneider, Teseo and Gil-Ureta, Francisca and Langlois, Timothy and Jiang, Chenfanfu and Zorin, Denis and Kaufman, Danny M. and Panozzo, Daniele},
title = {Intersection-free rigid body dynamics},
year = {2021},
issue_date = {August 2021},
publisher = {Association for Computing Machinery},
address = {New York, NY, USA},
volume = {40},
number = {4},
issn = {0730-0301},
url = {https://doi.org/10.1145/3450626.3459802},
doi = {10.1145/3450626.3459802},
journal = {ACM Trans. Graph.},
month = jul,
articleno = {183},
numpages = {16}
}

@article{Baumgarte1972,
title = {Stabilization of constraints and integrals of motion in dynamical systems},
journal = {Computer Methods in Applied Mechanics and Engineering},
volume = {1},
number = {1},
pages = {1-16},
year = {1972},
issn = {0045-7825},
doi = {https://doi.org/10.1016/0045-7825(72)90018-7},
url = {https://www.sciencedirect.com/science/article/pii/0045782572900187},
author = {J. Baumgarte}
}

@INPROCEEDINGS{Erez2015,
  author={Erez, Tom and Tassa, Yuval and Todorov, Emanuel},
  booktitle={2015 IEEE International Conference on Robotics and Automation (ICRA)}, 
  title={Simulation tools for model-based robotics: Comparison of Bullet, Havok, MuJoCo, ODE and PhysX}, 
  year={2015},
  volume={},
  number={},
  pages={4397-4404},
  doi={10.1109/ICRA.2015.7139807}}

@article{baraff1997introduction,
  title={An introduction to physically based modeling: rigid body simulation I—unconstrained rigid body dynamics},
  author={Baraff, David},
  journal={SIGGRAPH course notes},
  volume={82},
  year={1997}
}

@book{boyd2004convex,
  title={Convex optimization},
  author={Boyd, Stephen and Vandenberghe, Lieven},
  year={2004},
  publisher={Cambridge university press}
}

@incollection{kim2020dynamic,
  title={Dynamic deformables: implementation and production practicalities},
  author={Kim, Theodore and Eberle, David},
  booktitle={Acm siggraph 2020 courses},
  pages={1--182},
  year={2020}
}

@inproceedings{provot1995deformation,
  title={Deformation constraints in a mass-spring model to describe rigid cloth behaviour},
  author={Provot, Xavier and others},
  booktitle={Graphics interface},
  pages={147--147},
  year={1995},
  organization={Canadian Information Processing Society}
}

@book{press2007numerical,
  title={Numerical recipes 3rd edition: The art of scientific computing},
  author={Press, William H},
  year={2007},
  publisher={Cambridge university press}
}

@book{erleben2005physics,
  title={Physics-based animation},
  author={Erleben, Kenny and Sporring, Jon and Henriksen, Knud and Dohlmann, Henrik},
  volume={79},
  year={2005},
  publisher={Charles River Media Hingham}
}

@inproceedings{muller2004interactive,
  title={Interactive Virtual Materials.},
  author={M{\"u}ller, Matthias and Gross, Markus H},
  booktitle={Graphics interface},
  volume={2004},
  pages={239--246},
  year={2004}
}

@book{featherstone2008rigid,
  title={Rigid body dynamics algorithms},
  author={Featherstone, Roy},
  year={2008},
  publisher={Springer}
}

@inproceedings{todorov2012mujoco,
  title        = {MuJoCo: A physics engine for model-based control},
  author       = {Todorov, Emanuel and Erez, Tom and Tassa, Yuval},
  booktitle    = {2012 IEEE/RSJ International Conference on Intelligent Robots and Systems (IROS)},
  pages        = {5026--5033},
  year         = {2012},
  organization = {IEEE},
  doi          = {10.1109/IROS.2012.6386109}
}

@inproceedings{coumans2015bullet,
  title     = {Bullet Physics Simulation},
  author    = {Coumans, Erwin},
  booktitle = {ACM SIGGRAPH 2015 Courses},
  year      = {2015},
  publisher = {ACM},
  doi       = {10.1145/2776880.2792704}
}

@misc{nvidia2025physx,
  author       = {{NVIDIA Corporation}},
  title        = {NVIDIA PhysX SDK},
  howpublished = {\url{https://github.com/NVIDIA-Omniverse/PhysX}},
  year         = {2025},
  note         = {Accessed: 2026-01-04}
}

@inproceedings{DER,
author = {Bergou, Mikl\'{o}s and Wardetzky, Max and Robinson, Stephen and Audoly, Basile and Grinspun, Eitan},
title = {Discrete elastic rods},
year = {2008},
isbn = {9781450301121},
publisher = {Association for Computing Machinery},
address = {New York, NY, USA},
url = {https://doi.org/10.1145/1399504.1360662},
doi = {10.1145/1399504.1360662},
booktitle = {ACM SIGGRAPH 2008 Papers},
articleno = {63},
numpages = {12},
location = {Los Angeles, California},
series = {SIGGRAPH '08}
}

@article{Marsden_West_2001, title={Discrete mechanics and variational integrators}, volume={10}, DOI={10.1017/S096249290100006X}, journal={Acta Numerica}, author={Marsden, J. E. and West, M.}, year={2001}, pages={357–514}}

@book{Hairer2006,
  author    = {Hairer, Ernst and Lubich, Christian and Wanner, Gerhard},
  title     = {Geometric Numerical Integration: Structure-Preserving Algorithms for Ordinary Differential Equations},
  publisher = {Springer},
  edition   = {2},
  year      = {2006},
  doi       = {10.1007/3-540-30666-8}
}

@misc{eigenweb,
  title        = {Eigen: A C++ Template Library for Linear Algebra},
  author       = {Guennebaud, Ga{\"e}l and Jacob, Beno{\^i}t and others},
  howpublished = {\url{https://eigen.tuxfamily.org}},
  year         = {2010}
}

@incollection{wang2014intel,
  title={Intel math kernel library},
  author={Wang, Endong and Zhang, Qing and Shen, Bo and Zhang, Guangyong and Lu, Xiaowei and Wu, Qing and Wang, Yajuan},
  booktitle={High-Performance Computing on the Intel{\textregistered} Xeon Phi™: How to Fully Exploit MIC Architectures},
  pages={167--188},
  year={2014},
  publisher={Springer}
}

@book{goldstein1950classical,
  title={Classical mechanics},
  author={Goldstein, Herbert and Poole, Charles P and Safko, John},
  volume={2},
  year={1950},
  publisher={Addison-wesley Reading, MA}
}

@article{Yan2020ACE2,
  title={Structural basis for the recognition of SARS-CoV-2 by full-length human ACE2},
  author={Yan, Renhong and Zhang, Yuanyuan and Li, Yaning and Xia, Lu and Guo, Yingying and Zhou, Qiang},
  journal={Science},
  volume={367},
  number={6485},
  pages={1444--1448},
  year={2020},
  publisher={American Association for the Advancement of Science}
}

@article{Walls2020Spike,
  title={Structure, function, and antigenicity of the SARS-CoV-2 spike glycoprotein},
  author={Walls, Alexandra C and Park, Young-Jun and Tortorici, M Alejandra and Wall, Abigail and McGuire, Andrew T and Veesler, David},
  journal={Cell},
  volume={181},
  number={2},
  pages={281--292},
  year={2020},
  publisher={Elsevier}
}

\appendix

\end{document}